\documentclass[journal]{IEEEtran}
\usepackage{amsmath,amsfonts}
\usepackage{amssymb}
\usepackage{algorithmic}
\usepackage{algorithm}
\usepackage{array}
\usepackage[caption=false,font=footnotesize]{subfig}
\usepackage{textcomp}
\usepackage{stfloats}
\usepackage{url}
\usepackage{verbatim}
\usepackage{graphicx}
\usepackage{cite}
\usepackage{multirow}
\usepackage{xcolor}
\usepackage[nolist,nohyperlinks]{acronym} 
\newtheorem{remark}{Remark}

\begin{document}
\bstctlcite{BSTcontrol}

\title{Channel Estimation and Beamforming for Microwave Linear Analog Computers (MiLACs)-Aided Multiuser MISO Systems}

\author{Qiaosen Zhang, Matteo Nerini,~\IEEEmembership{Senior Member,~IEEE}, Bruno Clerckx,~\IEEEmembership{Fellow,~IEEE}
\thanks{Qiaosen Zhang, Matteo Nerini, and Bruno Clerckx are with the Department of Electrical and Electronic Engineering, Imperial College London, SW7 2AZ London, U.K (e-mail: qiaosen.zhang23@imperial.ac.uk; m.nerini20@imperial.ac.uk; b.clerckx@imperial.ac.uk).}
\thanks{}}

\markboth{}%
{}

\IEEEpubid{}

\maketitle

\begin{acronym}
    \acro{ADCs/DACs}{analog-to-digital/digital-to-analog converters}
    \acro{AoA}{angle of arrival}
    \acro{AS}{angular spread}
    \acro{ASD}{angular standard deviation}
    \acro{AWGN}{additive white Gaussian noise}
    \acro{BS}{base station}
    \acro{CSI}{channel state information}
    \acro{DFT}{Discrete Fourier Transform}
    \acro{DoF}{degree of freedom}
    \acro{EM}{electromagnetic}
    \acro{i.i.d.}{independent and identically distributed}
    \acro{JSDM}{joint spatial division and multiplexing}
    \acro{LS}{least squares}
    \acro{LMMSE}{linear minimum mean square error}
    \acro{MiLAC}{microwave linear analog computer}
    \acro{MSE}{mean square error}
    \acro{MIMO}{multiple-input multiple-output}
    \acro{MISO}{multiple-input single-output}
    \acro{MMSE}{minimum mean square error}
    \acro{MU}{multiuser}
    \acro{MU-MISO}{multiuser multiple-input single-output}
    \acro{NMSE}{normalized mean square error}
    \acro{RF}{radio-frequency}
    \acro{R-ZFBF}{regularized zero-forcing beamforming}
    \acro{RIS}{reconfigurable intelligent surface}
    \acro{SIM}{stacked intelligent metasurface}
    \acro{SNR}{signal-to-noise ratio}
    \acro{SVD}{singular value decomposition}
    \acro{TDD}{time division duplex}
    \acro{ULA}{uniform linear array}
    \acro{5G}{fifth-generation}
    \acro{6G}{sixth-generation}
\end{acronym}

\begin{abstract}
Microwave linear analog computers (MiLACs) have recently gained attention for future gigantic multiple-input multiple-output (MIMO) systems by enabling beamforming with greatly reduced hardware and computational cost. 
However, channel estimation for MiLAC-aided multiuser systems remains an open problem. 
Conventional channel estimation requires many radio-frequency (RF) chains to access full-dimensional received signals, followed by massive digital processing, which undermines the advantages of MiLAC-aided systems in reducing the number of RF chains and computational complexity.
In this paper, we propose computationally efficient channel estimation and beamforming schemes for MiLAC-aided multiuser multiple-input single-output (MU-MISO) systems with a limited number of RF chains.
We consider the general case where different user groups experience different channel correlation matrices.
By exploiting the rank deficiency of these matrices, the proposed schemes use MiLAC to compress the full-dimensional received signals in the analog domain, making them compatible with the available RF chains while preserving the essential channel information. 
Then, in the digital domain, only low-dimensional channel estimation is performed based on these compressed observations, substantially reducing computational cost.
We further show how regularized zero-forcing beamforming (R-ZFBF) can be efficiently realized from the low-dimensional channel estimates through a cascade of two MiLACs, which offers greater computational flexibility than a single MiLAC. 
Numerical results show that the proposed schemes reduce computational complexity up to $1540\times$ and $16108\times$, for channel estimation and beamforming, respectively, while achieving performance comparable to digital baselines.
\end{abstract}


\section{Introduction}

\IEEEPARstart{F}{u}ture wireless networks are expected to support extremely high data rates, ultra-low latency, and exceptional reliability. 
To meet these requirements, gigantic \ac{MIMO} with up to thousands of antennas has emerged as a promising paradigm beyond massive \ac{MIMO} \cite{Intro_1,Intro_2}.
By substantially enlarging the antenna array, gigantic \ac{MIMO} can provide enhanced spatial multiplexing, beamforming gains, and ultra-massive connectivity. 
However, such scaling makes conventional digital beamforming increasingly impractical. 
In particular, it requires one dedicated \ac{RF} chain per antenna and incurs computational complexity that scales rapidly with the array size, leading to prohibitive hardware cost, power consumption, and latency. 

To address this issue, recent research has pivoted toward alternative beamforming strategies that offload intensive computations from the digital domain to the analog domain. 
In this context, \ac{MiLAC} has been proposed in \cite{MiLAC_Part1,MiLAC_Part2}. 
A \ac{MiLAC} is a multiport microwave network composed of tunable admittance components that process microwave signals in the analog domain. 
Once the admittance components are configured, \ac{MiLAC} realizes the desired beamforming matrix. When an input signal is applied, \ac{MiLAC} directly generates the corresponding beamforming output through wave propagation inside the network, without requiring any digital matrix-vector multiplication.
As a result, \ac{MiLAC}-aided beamforming provides flexibility comparable to digital beamforming, while requiring only a minimal number of \ac{RF} chains, supporting the use of low-resolution \ac{ADCs/DACs}, and substantially reducing computational complexity.

Existing studies on \ac{MiLAC} have mainly focused on modelling \cite{MiLAC_Part1}, beamforming design \cite{MiLAC_Part2,MiLAC_Capcity,MiLAC_zheyu,MiLAC_Tianyu_MU_MISO,MiLAC_yiyang_wideband}, architecture design \cite{MiLAC_reduced_circuit_complexity,MiLAC_Xiaohua_Twolayer}, and prototyping \cite{MiLAC_Hybrid_Couplers}. 
Specifically, the concept and modelling of \ac{MiLAC} were first introduced in \cite{MiLAC_Part1}, followed by its application to computationally efficient beamforming in \cite{MiLAC_Part2}. 
By further imposing lossless and reciprocal constraints on \ac{MiLAC}, beamforming designs for point-to-point \ac{MIMO} systems to achieve capacity and for \ac{MU-MISO} systems to maximize sum rate were investigated in \cite{MiLAC_Capcity,MiLAC_zheyu,MiLAC_Tianyu_MU_MISO,MiLAC_yiyang_wideband}. 
Reduced-complexity and two-layer \ac{MiLAC} architectures were studied in \cite{MiLAC_reduced_circuit_complexity,MiLAC_Xiaohua_Twolayer}. 
In addition, a hardware prototype based on hybrid couplers and phase shifters was developed in \cite{MiLAC_Hybrid_Couplers}. 

Although the above-mentioned works \cite{MiLAC_Part2,MiLAC_Capcity,MiLAC_zheyu,MiLAC_Tianyu_MU_MISO,MiLAC_yiyang_wideband,MiLAC_reduced_circuit_complexity,MiLAC_Xiaohua_Twolayer} rely heavily on accurate \ac{CSI} to configure the tunable admittance components in \ac{MiLAC}, none of them addresses channel estimation for \ac{MiLAC}-aided systems. 
Only one recent work \cite{Qiaosen_MiLAC_CE} developed computationally efficient channel estimation schemes for \ac{MiLAC}-aided systems. 
However, it focused on point-to-point \ac{MIMO} systems and assumed that the number of transmit and receive \ac{RF} chains equals the number of transmit and receive antennas, respectively.
This represents an important first step, but remains limited in scope.

To fully exploit the potential of \ac{MiLAC}-aided beamforming in gigantic \ac{MIMO}, it is essential to acquire accurate \ac{CSI} for \ac{MU} systems while preserving the key advantages of \ac{MiLAC}, namely low computational complexity and a minimal number of \ac{RF} chains. 
However, this is non-trivial due to two key challenges.
\textit{First}, there is a mismatch between the operation that can be efficiently computed by \ac{MiLAC} and the operations required by conventional channel estimation. 
Specifically, \ac{MiLAC} is well suited for beamforming because, once configured, it can apply an instantaneous linear transformation to an input signal and generate the corresponding beamforming output directly in the analog domain. 
However, conventional channel estimation in \ac{MU} systems requires a different type of processing. 
The \ac{BS} first collects the received signals from all users at the $M$ \ac{BS} antennas over $T_u$ training symbol durations to form the received signal matrix $\mathbf{Y}\in\mathbb{C}^{M\times T_u}$. 
Then, the \ac{BS} post-multiplies $\mathbf{Y}$ by the conjugate of user-specific training sequences to isolate the corresponding user channels \cite[Section~3.1]{Massive_MIMO_Book}.
Therefore, conventional channel estimation requires both storing all received signals into $\mathbf{Y}$ and post-multiplying it by the training sequences. 
These operations are fundamentally different from the instantaneous linear transformation that can be efficiently computed by \ac{MiLAC}. 
Hence, it remains unclear how to exploit the computational capability of \ac{MiLAC} to perform channel estimation in \ac{MU} systems while preserving low computational complexity.
\textit{Second}, the limited number of \ac{RF} chains prevents the \ac{MiLAC}-aided \ac{BS} from directly acquiring the full-dimensional received signals.
Hence, channel estimation must rely on low-dimensional observations produced by \ac{MiLAC}, which makes accurate \ac{CSI} acquisition challenging.

To address these challenges, in this work, we investigate computationally efficient channel estimation for \ac{MiLAC}-aided \ac{MU-MISO} systems with a limited number of \ac{RF} chains, and further study the corresponding beamforming design based on the estimated channels. 
The main contributions are as follows.


\textit{First}, 
we present a \ac{MiLAC}-aided \ac{MU-MISO} system model integrating uplink training and downlink beamforming. 
The model employs only as many \ac{RF} chains as users, explicitly accounting for the limited number of \ac{RF} chains in \ac{MiLAC}-aided systems. 
It also incorporates different channel correlation matrices across user groups, leading to low-dimensional channel representations that underpin the proposed channel estimation and beamforming schemes.

\textit{Second}, 
we propose computationally efficient channel estimation schemes for the \ac{MiLAC}-aided \ac{MU-MISO} system. 
The key idea is to leverage the computational capability of \ac{MiLAC} to perform analog compression before digital channel estimation.
Specifically, in the analog domain, \ac{MiLAC} exploits the rank deficiency of the correlation matrices to compress the full-dimensional received signals into low-dimensional observations, thereby preserving essential channel information while matching the limited number of \ac{RF} chains. 
Then, in the digital domain, only low-dimensional channel representations are estimated from these compressed observations using \ac{LS} and \ac{MMSE} estimators, avoiding massive digital processing of full-dimensional received signals.
Analytical results show that the proposed \ac{LS} and \ac{MMSE} schemes achieve, respectively, lower and identical \ac{MSE} compared with their digital counterparts, while requiring greatly lower computational complexity, at the expense of extra training overhead.

\textit{Third}, 
we develop \ac{MiLAC}-aided \ac{R-ZFBF} schemes based on low-dimensional channel estimates. 
We show that a single \ac{MiLAC} cannot realize \ac{R-ZFBF} from such estimates with low computational complexity. 
To address this limitation, we propose a cascade \ac{MiLAC} architecture formed by connecting two \ac{MiLAC}s in series, which enables efficient implementation of \ac{R-ZFBF} from low-dimensional channel estimates. 
Compared with conventional digital \ac{R-ZFBF}, the proposed schemes are analytically shown to achieve the same sum rate with significantly reduced computational complexity. 
When uplink training overhead is taken into account, a minor loss in effective sum rate occurs.

\textit{Fourth}, 
we provide numerical results to validate the proposed schemes. 
The results confirm the predicted channel estimation performance and computational complexity reductions over digital \ac{LS} and \ac{MMSE}. 
They also show that \ac{MiLAC}-aided \ac{R-ZFBF} achieves the same sum rate as digital \ac{R-ZFBF} when training overhead is omitted, while greatly reducing computational complexity.
The loss of effective sum rate due to extra training overhead is further quantified.

\textit{Organization}: 
Section~\ref{Sec.2} presents the channel model, the signal model, and the transmission protocol of a \ac{MiLAC}-aided \ac{MU-MISO} system. 
Section~\ref{Sec.3} illustrates the proposed \ac{MiLAC}-aided channel estimation schemes.
Section~\ref{Sec.4} introduces the cascade \ac{MiLAC} and the \ac{MiLAC}-aided beamforming schemes based on estimated channels. 
Section~\ref{Sec.5} provides numerical results to validate the proposed schemes in comparison to digital baselines. 
Section~\ref{Sec.6} concludes this work. 

\textit{Notation}:
In this paper, $\mathbb{R}$ and $\mathbb{C}$ denote the sets of real and complex numbers, and $\mathbb{E}[\cdot]$ denotes expectation. 
The operators $(\cdot)^*$, $(\cdot)^T$, $(\cdot)^H$, and $(\cdot)^{-1}$ denote conjugate, transpose, Hermitian transpose, and inversion. 
For a scalar $a$, $|a|$ and $\lceil a\rceil$ denote its absolute value and ceiling function. 
For a vector $\mathbf{a}$, $[\mathbf{a}]_i$, $\|\mathbf{a}\|$, and $\mathrm{diag}(\mathbf{a})$ denote its $i$-th entry, $\ell_2$-norm, and the diagonal matrix formed from its entries. 
For a matrix $\mathbf{A}$, $[\mathbf{A}]_{i,j}$, $[\mathbf{A}]_{i,:}$, and $[\mathbf{A}]_{:,j}$ denote its $(i,j)$-th entry, $i$-th row, and $j$-th column, while $\mathrm{tr}(\mathbf{A})$, $\mathrm{rank}(\mathbf{A})$, $\mathrm{vec}(\mathbf{A})$, and $\|\mathbf{A}\|_F$ denote its trace, rank, vectorization, and Frobenius norm. 
The notation $\mathrm{blkdiag}(\mathbf{A},\ldots,\mathbf{B})$ denotes a block diagonal matrix with diagonal blocks from $\mathbf{A}$ to $\mathbf{B}$, and $[\mathbf{A}]_{\mathcal{I},\mathcal{J}}$ denotes the submatrix of $\mathbf{A}$ indexed by row set $\mathcal{I}$ and column set $\mathcal{J}$. 
For a set $\mathcal{I}$, $|\mathcal{I}|$ denotes its cardinality and $[\mathbf{a}_i]_{i\in\mathcal{I}}$ denotes the matrix formed by horizontally concatenating $\mathbf{a}_i$, $i\in\mathcal{I}$, all in the order induced by $\mathcal{I}$. 
Finally, $\mathbf{I}_N$ and $\mathbf{0}_{N\times M}$ denote the $N\times N$ identity matrix and the $N\times M$ all-zero matrix.

\section{System Model}
\label{Sec.2}
In this section, we present the uplink and downlink channel and signal models of a \ac{MiLAC}-aided \ac{MU-MISO} system, and describe the corresponding transmission protocol. 

\begin{figure*}[!t]
\centering
\subfloat[]{\includegraphics[width=0.85\columnwidth]{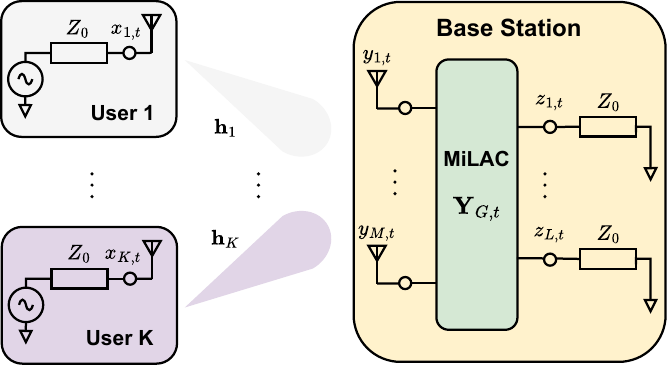}
\label{fig.1.a}}
\hfil
\subfloat[]{\includegraphics[width=0.85\columnwidth]{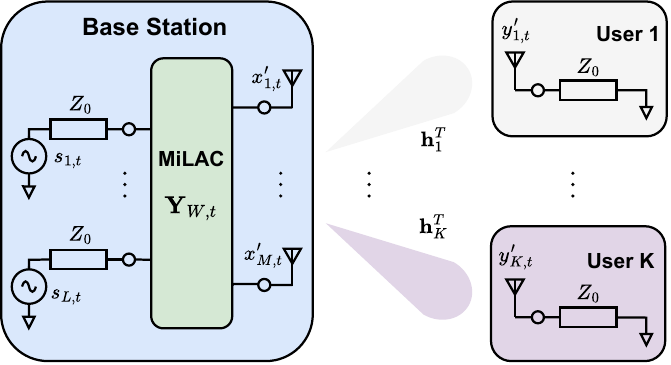}
\label{fig.1.b}}
\caption{A MiLAC-aided MU-MISO system: (a) uplink channel estimation; (b) downlink data transmission.}
\label{fig.1}
\end{figure*}

\subsection{Channel Model}
\label{Sec.2.1}
We consider a \ac{MU-MISO} system, where an $M$-antenna \ac{BS} aided by a \ac{MiLAC} serves $K$ single-antenna users under \ac{TDD} operation, as illustrated in Fig.~\ref{fig.1}. 

In the uplink, the channel from user $k$ to the \ac{BS} is denoted as $\mathbf{h}_k \in \mathbb{C}^{M \times 1}$, $\forall k \in \mathcal{K}=\{1,\dots,K\}$, and is modelled by $\mathbf{h}_k \sim\mathcal{CN}(\mathbf{0},\mathbf{R}_{k})$, where $\mathbf{R}_{k} = \mathbb{E}[\mathbf{h}_k \mathbf{h}_k^H]\in \mathbb{C}^{M \times M}$ denotes the channel correlation matrix. Using the Karhunen-Loeve expansion \cite[Section~2.2]{Massive_MIMO_Book}, the channel can be expressed as 
\begin{equation}
    \label{eq.SM.hk_raw}
    \mathbf{h}_k = \mathbf{U}_k\mathbf{\Lambda}_k^{\frac{1}{2}}\mathbf{v}_k,
\end{equation}
where $\mathbf{v}_k \in \mathbb{C}^{r_k \times 1}$ follows $\mathbf{v}_k \sim\mathcal{CN}(\mathbf{0},\mathbf{I}_{r_k})$, with $r_k = \mathrm{rank}(\mathbf{R}_k)$. The diagonal matrix $\mathbf{\Lambda}_k \in \mathbb{C}^{r_k \times r_k}$ collects the non-zero eigenvalues of $\mathbf{R}_k$ in descending order, while the semi-unitary matrix $\mathbf{U}_k \in \mathbb{C}^{M \times r_k}$ contains the corresponding eigenvectors, with $\mathbf{U}_k^H\mathbf{U}_k=\mathbf{I}_{r_k}$.

Equation \eqref{eq.SM.hk_raw} reveals that the channel of user $k \in \mathcal{K}$ lies in a low-dimensional subspace spanned by $\mathbf{U}_k$. In practice, users observed from the \ac{BS} with similar \ac{AoA} tend to exhibit highly overlapping subspaces \cite[Section~2.6]{Massive_MIMO_Book}. To exploit this overlap, user grouping has been widely adopted \cite{JSDM_1,JSDM_2,UE_grouping1,UE_grouping2}, enabling more efficient channel estimation and beamforming at the group level rather than individually.

In this work, we consider users partitioned into groups according to the similarity of their correlation matrices, following the principle of \ac{JSDM} \cite{JSDM_1,JSDM_2}. Let the $K$ users be partitioned into $G$ disjoint groups
\footnote{Unlike many existing works \cite{JSDM_1,JSDM_2,UE_grouping1,UE_grouping2}, which treat $G$ as a design parameter and study user clustering algorithms, we regard $G$ as an environmental property reflecting the spatial distribution of users. When $G=K$, the model reduces to the conventional setting without user grouping \cite{MiLAC_zheyu,MiLAC_Tianyu_MU_MISO}.}.
We denote the set of user indices in group $g$ by $\mathcal{K}_g$, $\forall g\in\mathcal G=\{1,\dots,G\}$. These sets satisfy $\bigcup_{g=1}^G \mathcal K_g=\mathcal K$ and $\mathcal K_g\cap\mathcal K_{g'}=\varnothing$ for $g\neq g'$. The number of users in group $g$ is denoted as $K_g=|\mathcal K_g|$, with $K=\sum_{g=1}^G K_g$. 

For analytical tractability, we assume that users within group $g$ share an identical correlation matrix $\mathbf{R}_g \in \mathbb{C}^{M\times M}$, while their large-scale fading coefficients $\beta_k$ may differ. Specifically, we model $\mathbf{R}_k = \beta_k \mathbf{R}_g$, $\forall k\in\mathcal K_g$. Let $\mathbf{\Lambda}_g \in \mathbb C^{r_g\times r_g}$ and $\mathbf{U}_g\in\mathbb C^{M\times r_g}$ contain the non-zero eigenvalues and the corresponding eigenvectors of $\mathbf{R}_g$, respectively, where $r_g=\mathrm{rank}(\mathbf {R}_g)$ and $\mathbf{U}_g^H\mathbf{U}_g=\mathbf{I}_{r_g}$. Substituting this model into \eqref{eq.SM.hk_raw}, the channel of user $k \in\mathcal K_g$ can be expressed as
\begin{equation}
    \label{eq.SM.hvk}
    \mathbf{h}_{k} = \sqrt{\beta_{k}}\mathbf{U}_g\mathbf{\Lambda}_g^{\frac{1}{2}}\mathbf{v}_{k} = \mathbf{U}_g\mathbf{h}_{v,k},
\end{equation}
where $\mathbf{h}_{v,k} \in \mathbb{C}^{r_g\times 1}$ denotes the virtual channel, i.e., the low-dimensional representation of $\mathbf{h}_k$ in the group subspace spanned by $\mathbf{U}_g$, and follows $\mathbf{h}_{v,k} \sim\mathcal{CN}(\mathbf{0},\mathbf{R}_{v,k})$, with $\mathbf{R}_{v,k} = \beta_k\mathbf{\Lambda}_g \in \mathbb{C}^{r_g \times r_g}$. This dimensionality reduction from $M$ to $r_g$, with $r_g\ll M$, motivates us to design uplink training in terms of the virtual channel, as detailed in Sections~\ref{Sec.3}.

Under \ac{TDD} operation, uplink and downlink channels are reciprocal. Hence, the downlink channel of user $k$ is given by the transpose of the uplink channel, i.e., $\mathbf{h}_k^T \in \mathbb{C}^{1\times M}$, $\forall k \in \mathcal{K}$. Consequently, the channel estimated from uplink training can be directly utilized for downlink beamforming.

\subsection{Signal Model}
\label{Sec.2.2}

Next, we present the uplink and downlink signal models for channel estimation and data transmission, respectively.

As shown in Fig.~\ref{fig.1.a}, in the uplink, the \ac{MiLAC}-aided \ac{BS} acts as an $M$-antenna receiver, equipped with $L$ RF chains, where $L\leq M$. Each \ac{RF} chain is modelled as a terminal loaded with a reference impedance $Z_0$ (typically $50~\Omega$) \cite{Nossek_Towards}. In this work, we set $L=K$, such that the number of RF chains matches the maximal data symbols transmitted in the downlink. Assume that each user $k \in \mathcal{K}$ transmits a training symbol $x_{k,t}$ with $|x_{k,t}|^2 = 1$ at symbol duration $t\in \mathcal{T}_u = \{1,\dots,T_u\}$, where $T_u$ denotes the number of symbol durations allocated to uplink channel estimation within a coherence block. The signal received at the \ac{BS} $\mathbf{y}_t \in \mathbb{C}^{M \times 1}$ is given by
\begin{equation}
    \label{eq.SM.yt_uplink}
    \mathbf{y}_t = \mathbf{H}\mathbf{P}\mathbf{x}_t+\mathbf{n}_t,
\end{equation}
where $\mathbf{x}_t = [x_{1,t},\dots,x_{K,t}]^T \in \mathbb{C}^{K\times 1}$, $\mathbf{H} = [\mathbf{h}_1,\dots,\mathbf{h}_K] \in \mathbb{C}^{M \times K}$, $\mathbf{P} = \mathrm{diag}(\sqrt{p_1},\dots,\sqrt{p_K})$, with $p_k$ denoting the transmit power of user $k$. The vector $\mathbf{n}_t \in \mathbb{C}^{M \times 1}$ denotes the \ac{AWGN}, following $\mathbf{n}_{t}\sim\mathcal{CN}(\mathbf{0},\sigma^2\mathbf{I}_{M})$, where $\sigma^2$ is the noise power. The signal $\mathbf{y}_t$ is compressed by the \ac{MiLAC}, giving the observation at the \ac{RF} chains $\mathbf{z}_t\in \mathbb{C}^{L \times 1}$ for channel estimation, written as 
\begin{equation}
    \label{eq.SM.zt_uplink}
    \mathbf{z}_t=\mathbf{G}_t\mathbf{y}_t,
\end{equation}
where $\mathbf{G}_t \in \mathbb{C}^{L \times M}$ denotes the training combiner implemented by the MiLAC. According to \cite{MiLAC_Part1,MiLAC_Part2}, the combiner $\mathbf{G}_t$ is a function of the admittance matrix of the MiLAC $\mathbf{Y}_{G,t} \in \mathbb{C}^{(M+L)\times(M+L)}$ as
\begin{equation}
\label{eq.SM.G_uplink}
    \mathbf{G}_t = \Bigl[\bigl(\frac{\mathbf{Y}_{G,t}}{Y_0}+\mathbf{I}_{M+L}\bigr)^{-1}\Bigr]_{M+\mathcal{L},\mathcal{M}} \;,
\end{equation}
where $Y_0 = 1/Z_0$ is the reference admittance, and $\mathcal{L} = \{1,\dots,L\}$ and $\mathcal{M} = \{1,\dots,M\}$ are introduced for brevity.

In contrast to the uplink, the \ac{MiLAC}-aided BS operates as an $M$-antenna transmitter in the downlink, as illustrated in Fig.~\ref{fig.1.b}. It transmits $L$ data symbols in parallel through $L$ RF chains, each of which is modelled as a voltage generator with a series impedance $Z_0$ \cite{Nossek_Towards}. For $t \in \mathcal{T}_d = \{1,\dots,T_d\}$, we denote the data symbol vector at the RF chains as $\mathbf{s}_t \in \mathbb{C}^{L \times 1}$, with $\mathbb{E}[\mathbf{s}_t \mathbf{s}_t^H] = (P_T/L)\,\mathbf{I}_{L}$, where $T_d$ is the number of symbol durations used for downlink data transmission within a coherence block and $P_T$ is the average transmit power. The vector $\mathbf{s}_t$ is precoded by the MiLAC, generating the signal transmitted at the \ac{BS} $\mathbf{x}'_t \in \mathbb{C}^{M \times 1}$, written as
\begin{equation}
    \label{eq.SM.xt'_downlink}
    \mathbf{x}'_t=\mathbf{W}_t\mathbf{s}_t,
\end{equation}
where $\mathbf{W}_t \in \mathbb{C}^{M\times L}$ denotes the data precoder implemented by the MiLAC. Referring to \cite{MiLAC_Part1,MiLAC_Part2}, the precoder $\mathbf{W}_t$ is a function of the admittance matrix of the MiLAC $\mathbf{Y}_{W,t} \in \mathbb{C}^{(L+M)\times(L+M)}$ as 
\begin{equation}
    \label{eq.SM.W_downlink}
    \mathbf{W}_t = \Bigl[\bigl(\frac{\mathbf{Y}_{W,t}}{Y_0}+\mathbf{I}_{L+M}\bigr)^{-1}\Bigr]_{L+\mathcal{M}, \mathcal{L}} \;.
\end{equation}
Finally, the signal received at user $k\in \mathcal{K}$ for symbol detection, is given by 
\begin{equation}
    \label{eq.SM.yk'_downlink}
    y'_{k,t} = \mathbf{h}_k^T\mathbf{x}'_t+n_{k,t},
\end{equation}
where $n_{k,t} \sim\mathcal{CN}(0,\sigma^2)$ is the \ac{AWGN}.

\begin{figure}[!t]
\centering
\includegraphics[width=0.95\columnwidth]{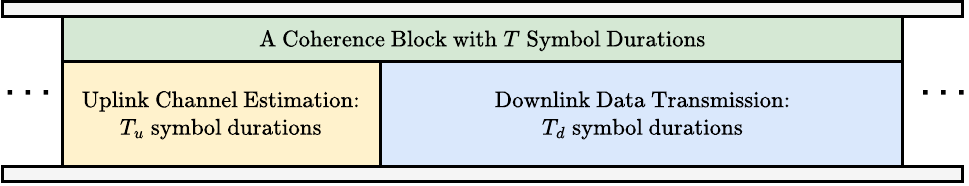}
\caption{A coherence block consisting of two phases with $T = T_u+T_d$ symbol durations.}
\label{fig.2}
\end{figure}
Overall, the signal models described above correspond to the two phases within each coherence block of the transmission protocol illustrated in Fig.~\ref{fig.2}, under the block-fading assumption that the channel between the users and the \ac{BS} remains approximately constant within each coherence block and varies independently across blocks. During the first $T_u$ symbol durations, users transmit training symbols to the \ac{BS} for uplink channel estimation. Specifically, the received signals $\mathbf{y}_t$ are first compressed by the training combiners $\mathbf{G}_t$ implemented by the \ac{MiLAC}, producing the observations $\mathbf{z}_t$, upon which digital processing estimates the \ac{CSI}, as detailed in Section~\ref{Sec.3}. Based on the acquired \ac{CSI}, the \ac{MiLAC} is then configured to implement the data precoders $\mathbf{W}_t$, enabling downlink data transmission over the remaining $T_d$ symbol durations within the coherence block, as discussed in Section~\ref{Sec.4}.

In this work, following \cite{MiLAC_Part1,MiLAC_Part2}, the admittance matrix of the \ac{MiLAC} is assumed to be fully reconfigurable, i.e., its entries can take arbitrary complex values 
\footnote{This assumption allows us to characterize the fundamental limits of \ac{MiLAC}-aided channel estimation and beamforming in terms of performance and computational complexity. Incorporating lossless and reciprocal constraints is left for future work.}. 
Consequently, the \ac{MiLAC} can realize any desired training combiner $\mathbf{G}_t$, $\forall t \in \mathcal{T}_u$, and data precoder $\mathbf{W}_t$, $\forall t \in \mathcal{T}_d$. 

\section{MiLAC-Aided Channel Estimation}
\label{Sec.3}
In this section, we propose computationally efficient channel estimation schemes for the \ac{MiLAC}-aided \ac{MU-MISO} system with a limited number of \ac{RF} chains.

\subsection{Training Protocol and Overall Observation Matrix}
\label{Sec.3.1}

\begin{figure}[!t]
\centering
\includegraphics[width=0.95\columnwidth]{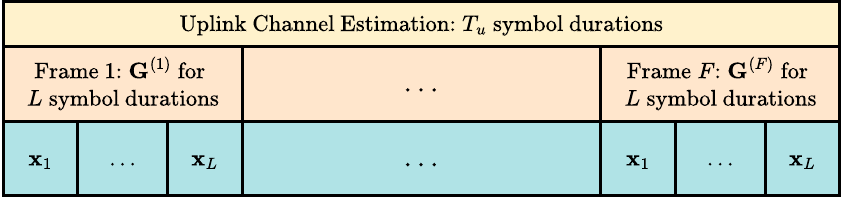}
\caption{Uplink channel estimation over $T_u$ symbol durations divided into $F$ frames of length $L$, where $T_u = FL$.}
\label{fig.4}
\end{figure}

With only $L$ \ac{RF} chains available, where $L=K$, the \ac{MiLAC}-aided \ac{BS} acquires the $L$-dimensional observation $\mathbf{z}_t$ at each symbol duration $t\in\mathcal{T}_u$, rather than the $M$-dimensional received signal $\mathbf{y}_t$.
To obtain sufficient observations for channel estimation under this RF-chain constraint, we first introduce a training protocol.

Specifically, following the training approach in \cite{Hybrid_CE_blockseperated,Hongyu_CE}, we divide the training phase of $T_u$ symbol durations into $F$ frames, each consisting of $L$ symbol durations, as illustrated in Fig.~\ref{fig.4}.
The value of $F$ is kept general here, which will be specified later according to the dimension of the channel representation to be estimated.
At each frame $f$, the training combiner $\mathbf{G}^{(f)}$ is kept fixed over the $L$ symbol durations, i.e., $\mathbf{G}_{(f-1)L+l} = \mathbf{G}^{(f)}$, $\forall f \in \mathcal{F} = \{1,\dots,F\}, l \in \mathcal{L} = \{1,\dots,L\}$, while it varies from frame to frame. 
Meanwhile, the same training matrix $\mathbf{X} = [\mathbf{x}_1,\dots,\mathbf{x}_L]^T \in \mathbb{C}^{L\times K}$ is reused at each frame, i.e., $\mathbf{x}_{(f-1)L+l} = \mathbf{x}_l, \forall f \in \mathcal{F}, l \in \mathcal{L}$. 
Here, the $l$-th row of $\mathbf{X}$, given by $\mathbf{x}_l^T$, contains the training symbols transmitted by the $K$ users at the $l$-th symbol duration of each frame, while the $k$-th column of $\mathbf{X}$, denoted by $\bar{\mathbf{x}}_k\in\mathbb{C}^{L\times1}$, represents the training sequence assigned to user $k \in \mathcal{K}$ over the $L$ symbol durations within each frame.
The matrix $\mathbf{X}$ is chosen as a \ac{DFT} matrix with $\mathbf{X}^H\mathbf{X}=L\mathbf{I}_K$, so that the user training sequences are mutually orthogonal and can be separated at the \ac{BS}.

Based on the above training protocol, we next collect the observations over the whole training phase into a compact matrix, which provides the starting point for the channel estimation design.
Specifically, within each frame $f \in \mathcal{F}$, we collect $\mathbf{z}_t$ over the $L$ symbol durations into $\mathbf{Z}^{(f)}=[\mathbf{z}_{(f-1)L+1},\dots,\mathbf{z}_{fL}]\in\mathbb{C}^{L\times L}$, which, by substituting \eqref{eq.SM.yt_uplink} into \eqref{eq.SM.zt_uplink}, can be written as
\begin{equation}
    \label{eq.CE.Zi}
    \mathbf{Z}^{(f)} = \mathbf{G}^{(f)}\mathbf{H}\mathbf{P}\mathbf{X}^T+\mathbf{G}^{(f)}\mathbf{N}^{(f)},
\end{equation}
where $\mathbf{N}^{(f)} = [\mathbf{n}_{(f-1)L+1},\dots,\mathbf{n}_{fL}]\in \mathbb{C}^{M\times L}$. Stacking \eqref{eq.CE.Zi} from all $F$ frames then gives the overall observation matrix $\mathbf{Z}\in \mathbb{C}^{FL\times L}$ as
\begin{equation}
    \label{eq.CE.Z_original}
    \mathbf{Z} = 
        \begin{bmatrix}
        \mathbf{Z}^{(1)}\\
        \vdots \\
        \mathbf{Z}^{(F)}\\
    \end{bmatrix}
    =    
    \underbrace{
        \begin{bmatrix}
            \mathbf{G}^{(1)}\\
            \vdots \\
            \mathbf{G}^{(F)}\\
        \end{bmatrix}
        }_{\mathbf{G} \in \mathbb{C}^{FL\times M}}
    \mathbf{H}\mathbf{P}\mathbf{X}^T 
    + 
    \underbrace{
        \begin{bmatrix}
            \mathbf{G}^{(1)}\mathbf{N}^{(1)}\\
            \vdots \\
            \mathbf{G}^{(F)}\mathbf{N}^{(F)}\\
        \end{bmatrix}
     }_{\mathbf{N} \in \mathbb{C}^{FL\times L}}
    ,
\end{equation}
where $\mathbf{G}$ denotes the equivalent training combiner implemented by the MiLAC over the whole training phase.
In the following subsections, we specify the low-dimensional representation of $\mathbf{H}$ and design $\mathbf{G}$ accordingly, such that $\mathbf{Z}$ preserves essential channel information while removing redundant dimensions. 
This allows the low-dimensional channel representation to be estimated efficiently in the digital domain.

\subsection{Virtual Channel Estimation}
\label{Sec.3.2}

We start with the virtual channel representation introduced in Section~\ref{Sec.2.1}.
Recall that the $K$ users are partitioned into $G$ groups, where group $g$ contains $K_g$ users, $\forall g \in \mathcal{G}$. 
From $\mathbf{h}_{k} = \mathbf{U}_g\mathbf{h}_{v,k}$, the channel of user $k \in \mathcal{K}_g$ lies in the low-dimensional group subspace spanned by $\mathbf{U}_g$, where $\mathbf{U}_g^H\mathbf{U}_g = \mathbf{I}_{r_g}$. 
Assuming $\mathbf{U}_g$ is known, it suffices to estimate the virtual channel $\mathbf{h}_{v,k}$ instead of the full-dimensional channel $\mathbf{h}_k$, thereby reducing the number of unknown channel parameters from $M$ to $r_g$. 

To estimate the virtual channels of all users from the overall observation matrix $\mathbf{Z}$, we first rewrite the channel matrix $\mathbf{H} = [\mathbf{h}_1,\dots,\mathbf{h}_K]$ in terms of these virtual channels.
From $\mathbf{h}_{k} = \mathbf{U}_g\mathbf{h}_{v,k}$, it follows that
\begin{equation}
    \label{eq.CE.H_intermsof_Hv}
    \mathbf{H} = \mathbf{U}\mathbf{H}_v,
\end{equation}
where $\mathbf{U} = [\mathbf{U}_1,\dots,\mathbf{U}_G] \in \mathbb{C}^{M\times r}$ with $r = \sum_{g \in \mathcal{G}}r_g$, and $\mathbf{H}_v = \mathrm{blkdiag}(\mathbf{H}_{v,1},\dots,\mathbf{H}_{v,G}) \in \mathbb{C}^{r\times K}$, in which $\mathbf{H}_{v,g} = [\mathbf{h}_{v,k}]_{k \in \mathcal{K}_g} \in \mathbb{C}^{r_g \times K_g}$. 
Since $\mathbf{H}_v$ has $r$ rows, the number of frames in the training protocol is set as $F=\lceil r/L\rceil$.
For analytical convenience, we consider $FL = r$ 
\footnote{If $FL>r$, the last $(FL-r)$ rows of $\mathbf{G}^{(F)}$ can be set to zero, which will not affect the proposed channel estimation scheme.}.

Given the representation of $\mathbf{H}$ in \eqref{eq.CE.H_intermsof_Hv}, we next design the equivalent training combiner $\mathbf{G}$. For group $g \in \mathcal{G}$, its channel matrix is $\mathbf{U}_g\mathbf{H}_{v,g}$.
Since $\mathbf{U}_g^H\mathbf{U}_g=\mathbf{I}_{r_g}$, applying $\mathbf{U}_g^H$ to $\mathbf{U}_g\mathbf{H}_{v,g}$ directly yields $\mathbf{H}_{v,g}$.
This motivates using $\mathbf{U}_g^H$ as the training combiner for group $g$.
By stacking the combiners of all groups, we set $\mathbf{G}=\mathbf{U}^H$.
Under this choice of $\mathbf{G}$, substituting \eqref{eq.CE.H_intermsof_Hv} into \eqref{eq.CE.Z_original}, the matrix $\mathbf{Z}$ can be repartitioned row-wise according to the dimensions $r_g$, $\forall g\in\mathcal{G}$, as
\begin{equation}
    \label{eq.CE.Z_G=U^H}
    \mathbf{Z} = 
    \begin{bmatrix}
        \mathbf{Z}_1\\
        \vdots \\
        \mathbf{Z}_G\\
    \end{bmatrix}
    =
    \begin{bmatrix}
        \mathbf{U}_1^H\\
        \vdots \\
        \mathbf{U}_G^H\\
    \end{bmatrix}
    \mathbf{U}\mathbf{H}_v\mathbf{P}\mathbf{X}^T
    +
    \begin{bmatrix}
        \mathbf{N}_1\\
        \vdots \\
        \mathbf{N}_G\\
    \end{bmatrix}
    ,
\end{equation}
where $\mathbf{Z}_g \in \mathbb{C}^{r_g \times L}$ and $\mathbf{N}_g \in \mathbb{C}^{r_g \times L}$ denote the observation and noise matrices of group $g$, respectively, with $\mathrm{vec}(\mathbf{N}_g) \sim \mathcal{CN}(\mathbf{0},\sigma^2\mathbf{I}_{r_gL})$ (see Appendix~\ref{Ap.noise}). 
The matrix $\mathbf{Z}_g$ is therefore given by
\begin{equation}
    \label{eq.CE.Z_g}
    \mathbf{Z}_g  = \mathbf{U}_g^H\mathbf{U}\mathbf{H}_v\mathbf{P}\mathbf{X}^T+\mathbf{N}_g, 
\end{equation}
where the projection by $\mathbf{U}_g^H$ exposes $\mathbf{H}_{v,g}$ for subsequent estimation.

Based on \eqref{eq.CE.Z_g}, virtual channel estimation can now be performed separately for each group. 
Specifically, for user $k \in \mathcal{K}_g$ in group $g \in \mathcal{G}$, we multiply $\mathbf{Z}_g$ by the conjugate of the training sequence $\bar{\mathbf{x}}_k^*$. 
Since $\mathbf{X}^H\mathbf{X}=L\mathbf{I}_K$, the effective observation $\bar{\mathbf{z}}_k \in \mathbb{C}^{r_g \times 1}$ for user $k$ is given by
\begin{equation}
    \label{eq.CE.bar_z_k}
    \bar{\mathbf{z}}_k = \mathbf{Z}_g\bar{\mathbf{x}}_k^* =  \sqrt{p_k}L\mathbf{h}_{v,k}+\bar{\mathbf{n}}_k,
\end{equation}
where $\bar{\mathbf{n}}_k = \mathbf N_g\bar{\mathbf{x}}_k^* \in \mathbb{C}^{r_g\times 1}$, with $\bar{\mathbf{n}}_k \sim \mathcal{CN}(\mathbf{0},\sigma^2L\mathbf{I}_{r_g})$. 
Without requiring any prior knowledge of channel second-order statistics
\footnote{The \ac{LS} estimator itself does not require prior channel statistics. 
However, to obtain $\mathbf{Z}_g$ and the resulting $\bar{\mathbf{z}}_k$, $\forall g \in \mathcal{G}, k \in \mathcal{K}_g$, the matrix $\mathbf U$ must be known for configuring the training combiners $\mathbf G^{(f)}$, $\forall f \in\mathcal{F}$.},
the \ac{LS} estimator of $\mathbf{h}_{v,k}$ is 
\begin{equation}
    \label{eq.CE.hvk_LS}
    \mathbf{h}_{v,k}^{\text{LS}} = \frac{1}{\sqrt{p_k}L}\bar{\mathbf{z}}_k,
\end{equation}
with the corresponding \ac{MSE} given by 
\begin{equation}
    \label{eq.CE.hvk_LS_MSE}
    \mathbb{E}\{\|\mathbf{h}_{v,k}-\mathbf{h}_{v,k}^{\text{LS}}\|^2\} = \mathbb{E}\left\{\left\|\frac{\bar{\mathbf{n}}_k}{\sqrt{p_k}L}\right\|^2\right\} = \frac{\sigma^2r_g}{p_kL}.
\end{equation}
By further exploiting the channel statistics, i.e., assuming $\mathbf{R}_k=\beta_k\mathbf{R}_g$ is available, the MMSE estimator of $\mathbf{h}_{v,k}$ is
\begin{equation}
    \label{eq.CE.hvk_MMSE}
    \mathbf{h}_{v,k}^{\text{MMSE}} = \mathbf{A}_k\bar{\mathbf{z}}_k,
\end{equation}
where $\mathbf{A}_k \in \mathbb{C}^{r_g \times r_g}$ is given by 
\begin{equation}
    \label{eq.CE.hvk_MMSE_Ak}
    \mathbf{A}_k = \sqrt{p_k}\mathbf{R}_{v,k}(p_kL\mathbf{R}_{v,k}+\sigma^2\mathbf{I}_{r_g})^{-1}, 
\end{equation}
which is diagonal since $\mathbf{R}_{v,k} = \beta_k\mathbf{\Lambda}_g$. The resulting \ac{MSE} is 
\begin{align}
    & \mathbb{E}\{\|\mathbf{h}_{v,k}-\mathbf{h}_{v,k}^{\text{MMSE}}\|^2\}\vphantom{\frac{p_kL}{\sigma^2}}\\
    & = \mathrm{tr}\Bigl(\mathbf{R}_{v,k} - p_kL\mathbf{R}_{v,k}\bigl(p_kL\mathbf{R}_{v,k}+\sigma^2\mathbf{I}_{r_g}\bigr)^{-1}\mathbf{R}_{v,k}\Bigr)\vphantom{\frac{p_kL}{\sigma^2}} \label{eq.CE.hvk_MMSE_MSE_full}\\
    & = \mathrm{tr}\Big(\big(\mathbf{R}_{v,k}^{-1}+\frac{p_kL}{\sigma^2}\mathbf{I}_{r_g}\big)^{-1} \Big)     \label{eq.CE.hvk_MMSE_MSE}.
\end{align}
Finally, collecting \eqref{eq.CE.hvk_LS} and \eqref{eq.CE.hvk_MMSE} over all $K$ users yields the estimated virtual channel matrices $\mathbf{H}_v^{\text{LS}}$ and $\mathbf{H}_v^{\text{MMSE}}$.

The scheme proposed in this subsection exploits the \ac{MiLAC} to perform group-wise subspace projection in the analog domain, thereby reducing digital estimation from full-dimensional channels to low-dimensional virtual channels. 
This scheme is particularly efficient when $\mathbf{U} = [\mathbf{U}_1,\dots,\mathbf{U}_G]$ has full column rank, i.e., when the group subspaces do not overlap, which typically holds for a small number of groups $G$.
However, as $G$ increases, overlap among group subspaces becomes more likely, causing $\mathbf{U}$ to lose full column rank. 
In this case, the aggregate dimension $r=\sum_{g\in\mathcal{G}}r_g$ exceeds $\mathrm{rank}(\mathbf{U})$ and may even exceed the number of \ac{BS} antennas $M$.
As a result, redundant channel parameters associated with overlapping subspaces are estimated, leading to excessive training overhead.
Moreover, the loss of full column rank of $\mathbf{U}$ affects the subsequent \ac{MiLAC}-aided beamforming design, since efficient precoder construction based on $\mathbf{H}_v$ relies on this property, as detailed in Section~\ref{Sec.4.3}.
To accommodate the case with large $G$, the global virtual channel estimation scheme is introduced in Section~\ref{Sec.3.3}.

\begin{remark}
    \label{Remark 1}  
    Since the correlation matrices $\mathbf{R}_k$, $\forall k \in \mathcal{K}$, vary slowly across coherence blocks \cite{SU_Keyref_Sayeed,SU_Keyref_Hybrid}, we assume they are known. 
    Hence, the matrix $\mathbf{U}$ and the admittance components of the MiLAC realizing the desired training combiners $\mathbf{G}^{(f)}$, $\forall f \in \mathcal{F}$, can be obtained offline.
    Using the definition of computational complexity in \cite[Appendix]{MiLAC_Part1}, the complexity of the proposed scheme is dominated by the matrix-vector product $\mathbf{Z}_g\bar{\mathbf{x}}_k^*$ used to obtain $\bar{\mathbf{z}}_k$ for all $K$ users, which requires $8L\sum_{g \in \mathcal{G}} r_gK_g$ real operations. 
    Since the subsequent LS and MMSE estimators involve only entry-wise scaling of $\bar{\mathbf{z}}_k$, their complexity is negligible. 
    Assuming equal group sizes, i.e., $K_g=K/G$, and recalling that $L=K$, the complexity of the proposed scheme simplifies to $8rK^2/G$ for both estimators.
\end{remark}

\subsection{Global Virtual Channel Estimation}
\label{Sec.3.3}
We now introduce the global virtual channel estimation scheme to accommodate the case with a large number of groups $G$, where overlap among group subspaces arises. 
The key idea is to obtain a full-column-rank matrix that spans the same subspace as $\mathbf{U}=[\mathbf{U}_1,\dots,\mathbf{U}_G]$, and use it to construct a low-dimensional representation of the channel matrix $\mathbf{H}$.
To this end, we take the \ac{SVD} of $\mathbf{U}$ as
\begin{equation}
    \label{eq.CE.U_SVD}
    \mathbf{U} = \tilde{\mathbf{U}}\mathbf{\Sigma}\mathbf{V}^H,
\end{equation}
where the diagonal matrix $\mathbf{\Sigma} \in \mathbb{C}^{\tilde{r} \times \tilde{r}}$ contains the non-zero singular values of $\mathbf{U}$ in descending order, while the semi-unitary matrices $\tilde{\mathbf{U}} \in \mathbb{C}^{M \times \tilde{r}}$ and $\mathbf{V} \in \mathbb{C}^{r \times \tilde{r}}$ contain the corresponding left and right singular vectors. 
Here, $\tilde{r} = \mathrm{rank}(\mathbf{U})\leq\min\{M,r\}$ and $\tilde{\mathbf{U}}^H\tilde{\mathbf{U}} = \mathbf{V}^H\mathbf{V} = \mathbf{I}_{\tilde{r}}$.
By substituting \eqref{eq.CE.U_SVD} into \eqref{eq.CE.H_intermsof_Hv}, the channel $\mathbf{H}$ is re-expressed as 
\begin{equation}
    \label{eq.CE.H_intermsof_Hvtilde}
    \mathbf{H} = \tilde{\mathbf{U}}\tilde{\mathbf{H}}_v,
\end{equation}
where $\tilde{\mathbf{H}}_v = \mathbf{\Sigma}\mathbf{V}^H\mathbf{H}_v \in \mathbb{C}^{\tilde{r} \times K}$ denotes the low-dimensional representation of $\mathbf{H}$ in the global subspace spanned by $\tilde{\mathbf{U}}$, referred to as the global virtual channel. 
Equivalently, $\tilde{\mathbf{H}}_v = [\tilde{\mathbf{h}}_{v,1},\dots,\tilde{\mathbf{h}}_{v,K}]$, where $\tilde{\mathbf{h}}_{v,k} \sim\mathcal{CN}(\mathbf{0},\tilde{\mathbf{R}}_{v,k})$, with $\tilde{\mathbf{R}}_{v,k} = \tilde{\mathbf{U}}^H\mathbf{R}_k\tilde{\mathbf{U}} \in \mathbb{C}^{\tilde{r} \times \tilde{r}}$, $\forall k \in \mathcal{K}$.

Given the representation of $\mathbf{H}$ in \eqref{eq.CE.H_intermsof_Hvtilde}, we 
follow the training protocol in Section~\ref{Sec.3.1} to estimate $\tilde{\mathbf{H}}_v$ from the overall observation matrix $\mathbf{Z}$ in \eqref{eq.CE.Z_original}. 
Since $\tilde{\mathbf{H}}_v$ has $\tilde{r}$ rows, the number of frames is set as $F=\lceil \tilde{r}/L\rceil$.
For analytical convenience, we consider $FL=\tilde{r}$.
Substituting \eqref{eq.CE.H_intermsof_Hvtilde} into \eqref{eq.CE.Z_original}, we have
\begin{equation}
    \label{eq.CE.Z_original_Hvtilde}
    \mathbf{Z} = \mathbf{G}\tilde{\mathbf{U}}\tilde{\mathbf{H}}_v\mathbf{P}\mathbf{X}^T+
    \mathbf{N}.
\end{equation}
Unlike the virtual channel matrix $\mathbf{H}_v$ in Section~\ref{Sec.3.2}, which is block diagonal and can be estimated group by group, the matrix $\tilde{\mathbf{H}}_v$ does not have such a structure, with each of its column representing the global virtual channel of one user.
Thus, we design $\mathbf{G}$ such that $\tilde{\mathbf{H}}_v$ is directly exposed in $\mathbf{Z}$.
Since $\tilde{\mathbf{U}}^H\tilde{\mathbf{U}}=\mathbf{I}_{\tilde r}$, we set $\mathbf{G}=\tilde{\mathbf{U}}^H$, so that \eqref{eq.CE.Z_original_Hvtilde} becomes
\begin{equation}
    \label{eq.CE.Z_intermsof_Hvtilde}
    \mathbf{Z} = \tilde{\mathbf{H}}_v\mathbf{P}\mathbf{X}^T+\mathbf{N},
\end{equation}
with $\mathrm{vec}(\mathbf{N}) \sim \mathcal{CN}(\mathbf{0},\sigma^2\mathbf{I}_{\tilde{r}L})$, which follows from the same argument in Appendix~\ref{Ap.noise} since the rows of $\mathbf{G}$ are orthonormal.

Based on \eqref{eq.CE.Z_intermsof_Hvtilde}, we can now perform global virtual channel estimation for all $K$ users. 
Specifically, for user $k \in \mathcal{K}$, we multiply $\mathbf{Z}$ by the conjugate of the training sequence $\bar{\mathbf{x}}_k^*$, yielding the effective observation $\tilde{\mathbf{z}}_k \in \mathbb{C}^{\tilde{r}\times 1}$ as
\begin{equation}
    \label{eq.CE.tilde_bar_z_k}
    \tilde{\mathbf{z}}_k = \mathbf{Z}\bar{\mathbf{x}}_k^* = \sqrt{p_k}L\tilde{\mathbf{h}}_{v,k}+\tilde{\mathbf{n}}_k
\end{equation}
where $\tilde{\mathbf{n}}_k = \mathbf{N}\bar{\mathbf{x}}_k^* \in \mathbb{C}^{\tilde{r}\times 1}$ follows $\tilde{\mathbf{n}}_k \sim \mathcal{CN}(\mathbf{0},\sigma^2L\mathbf{I}_{\tilde{r}})$. 
Hence, the \ac{LS} estimator of $\tilde{\mathbf{h}}_{v,k}$ is
\begin{equation}
    \label{eq.CE.hvktilde_LS}
    \tilde{\mathbf{h}}_{v,k}^{\text{LS}} = \frac{1}{\sqrt{p_k}L}\tilde{\mathbf{z}}_k,
\end{equation}
with the corresponding \ac{MSE} given by
\begin{equation}
    \label{eq.CE.hvktilde_LS_MSE}
    \mathbb{E}\{\|\tilde{\mathbf{h}}_{v,k}-\tilde{\mathbf{h}}_{v,k}^{\text{LS}}\|^2\} = \mathbb{E}\left\{\left\|\frac{\tilde{\mathbf{n}}_k}{\sqrt{p_k}L}\right\|^2\right\} = \frac{\sigma^2\tilde{r}}{p_kL}.
\end{equation}
Assuming the correlation matrix $\mathbf{R}_k$ is known, the MMSE estimator of $\tilde{\mathbf{h}}_{v,k}$ is 
\begin{equation}
    \label{eq.CE.hvktilde_MMSE}
    \tilde{\mathbf{h}}_{v,k}^{\text{MMSE}} = \tilde{\mathbf{A}}_k\tilde{\mathbf{z}}_k,
\end{equation}
where $\tilde{\mathbf{A}}_k \in \mathbb{C}^{\tilde{r} \times \tilde{r}}$ is given by 
\begin{equation}
    \label{eq.CE.hvktilde_MMSE_Ak}
    \tilde{\mathbf{A}}_k = \sqrt{p_k}\tilde{\mathbf{R}}_{v,k}(p_kL\tilde{\mathbf{R}}_{v,k}+\sigma^2\mathbf{I}_{\tilde{r}})^{-1}.
\end{equation}
The resulting \ac{MSE} is 
\begin{align}
    & \mathbb{E}\{\|\tilde{\mathbf{h}}_{v,k}-\tilde{\mathbf{h}}_{v,k}^{\text{MMSE}}\|^2\}\\
    & = \mathrm{tr}\Bigl(\tilde{\mathbf{R}}_{v,k} - p_kL\tilde{\mathbf{R}}_{v,k}\bigl(p_kL\tilde{\mathbf{R}}_{v,k}+\sigma^2\mathbf{I}_{\tilde{r}}\bigr)^{-1}\tilde{\mathbf{R}}_{v,k}\Bigr)     \label{eq.CE.hvktilde_MMSE_MSE}.
\end{align}
Finally, the estimated global virtual channel matrices $\tilde{\mathbf{H}}_v^{\text{LS}}$ and $\tilde{\mathbf{H}}_v^{\text{MMSE}}$ are obtained by collecting \eqref{eq.CE.hvktilde_LS} and \eqref{eq.CE.hvktilde_MMSE} of all $K$ users.

\begin{remark}
    \label{Remark 2}
    Given the correlation matrices $\mathbf{R}_k$, $\forall k \in \mathcal{K}$, the matrix $\tilde{\mathbf{U}}$ and the admittance components of the MiLAC realizing the desired training combiners $\mathbf{G}^{(f)}$, $\forall f\in\mathcal{F}$, can be precomputed offline. 
    Thus, the computational complexity of the proposed scheme only arises from two digital operations. 
    The first is the matrix-vector product $\mathbf{Z}\bar{\mathbf{x}}_k^*$ used to obtain $\tilde{\mathbf{z}}_k$ for all $K$ users, requiring $8\tilde{r} KL$ real operations. 
    The second is applying the LS and MMSE estimators. 
    The LS estimator incurs negligible cost since it only involves entry-wise scaling of $\tilde{\mathbf{z}}_k$. 
    In contrast, the MMSE estimator requires multiplying $\tilde{\mathbf{z}}_k$ by $\tilde{\mathbf{A}}_k$. 
    Although $\tilde{\mathbf{A}}_k$ can be precomputed offline given $\mathbf{R}_k$, applying it to all $K$ users still requires $8\tilde{r}^2K$ real operations.   
    Overall, recalling that $L = K$, the complexity of the proposed scheme using the LS estimator and the MMSE estimator is $8\tilde{r}K^2$ and $8(\tilde{r}K^2+\tilde{r}^2K)$ real operations, respectively.
\end{remark}

\subsection{Comparison with Conventional Digital Channel Estimation}
\label{Sec.3.4}

So far, we have introduced two MiLAC-aided channel estimation schemes: the virtual channel estimation scheme in Section~\ref{Sec.3.2} for small $G$, and the global virtual channel estimation scheme in Section~\ref{Sec.3.3} for large $G$, where overlap among group subspaces arises. 
In this subsection, we compare them with conventional digital channel estimation.

\begin{figure}[!t]
\centering
\includegraphics[width=0.75\columnwidth]{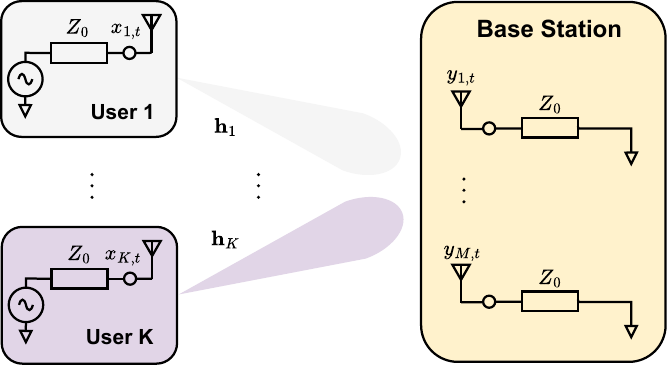}
\caption{Uplink channel estimation in a digital \ac{MU-MISO} system}
\label{fig.5}
\end{figure}
To facilitate the comparison, we first revisit conventional LS and MMSE channel estimation in a digital MU-MISO system, as illustrated in Fig.~\ref{fig.5}. Unlike the MiLAC-aided system in Fig.~\ref{fig.1.a}, no analog projection is performed and the BS directly observes the full-dimensional received signal $\mathbf{y}_t$ in \eqref{eq.SM.yt_uplink}, which requires one RF chain per BS antenna, i.e., $L=M$. To estimate $KM$ unknown entries in the channel $\mathbf{H}$ with minimal training overhead 
\footnote{The digital system estimates $\mathbf{H}$ rather than $\mathbf{H}_v$ or $\tilde{\mathbf{H}}_v$, since obtaining the latter digitally would incur higher computational complexity in general.},
we set $T_u=K$. Let each user $k\in\mathcal{K}$ transmit a training sequence $\bar{\mathbf{x}}_{k} \in\mathbb{C}^{T_u\times 1}$ over $T_u$ symbol durations, where $\bar{\mathbf{x}}_{k}$ is the $k$-th column of the training matrix $\mathbf{X}=[\bar{\mathbf{x}}_{1},\dots,\bar{\mathbf{x}}_{K}]\in\mathbb{C}^{T_u\times K}$ satisfying $\mathbf{X}^H\mathbf{X}=T_u\mathbf{I}_K$. By collecting $\mathbf{y}_t$ over $T_u$ symbol durations, we obtain $\mathbf{Y} = [\mathbf{y}_1,\dots,\mathbf{y}_{T_u}] \in \mathbb{C}^{M\times T_u}$, written as 
\begin{equation}
    \label{eq.CE.digital_Y}
    \mathbf{Y} = \mathbf{H}\mathbf{P}\mathbf{X}^T+\mathbf{N},
\end{equation}
where $\mathbf{N} = [\mathbf{n}_1,\dots,\mathbf{n}_{T_u}] \in \mathbb{C}^{M \times T_u}$ satisfies $\mathrm{vec}(\mathbf{N}) \sim \mathcal{CN}(\mathbf{0},\sigma^2\mathbf{I}_{T_uM})$.

Based on \eqref{eq.CE.digital_Y}, channel estimation can be carried out for all $K$ users. Specifically, for user $k \in \mathcal{K}$, multiplying $\mathbf{Y}$ by the conjugate of the training sequence $\bar{\mathbf{x}}_{k}^*$ yields the effective observation $\mathbf{z}_{d,k} \in \mathbb{C}^{M\times 1}$ as 
\begin{equation}
    \label{eq.CE.digital_zdk}
    \mathbf{z}_{d,k} = \mathbf{Y}\bar{\mathbf{x}}_{k}^* = \sqrt{p_k}T_u\mathbf{h}_k+\mathbf{n}_{d,k},
\end{equation}
where $\mathbf{n}_{d,k} = \mathbf{N}\bar{\mathbf{x}}_{k}^* \in \mathbb{C}^{M\times 1}$, with $\mathbf{n}_{d,k} \sim \mathcal{CN}(\mathbf{0},\sigma^2T_u\mathbf{I}_{M})$. 
The LS estimator of $\mathbf{h}_k$ is then given by 
\begin{equation}
    \label{eq.CE.digital_hk_LS}
    \mathbf{h}_k^{\text{LS}} = \frac{1}{\sqrt{p_k}T_u}\mathbf{z}_{d,k},
\end{equation}
with the corresponding \ac{MSE} given by 
\begin{equation}
    \label{eq.CE.digital_hk_LS_MSE}
    \mathbb{E}\{\|\mathbf{h}_{k}-\mathbf{h}_k^{\text{LS}} \|^2\} = \mathbb{E}\left\{\left\|\frac{\mathbf{n}_{d,k}}{\sqrt{p_k}T_u}\right\|^2\right\} = \frac{\sigma^2M}{p_kT_u}.
\end{equation}
When the correlation matrix $\mathbf{R}_k$ is available, the MMSE estimator of $\mathbf{h}_{k}$ is 
\begin{equation}
    \label{eq.CE.digital_hk_MMSE}
    \mathbf{h}_k^{\text{MMSE}} = \mathbf{A}_{d,k}\mathbf{z}_{d,k},
\end{equation}
where $\mathbf{A}_{d,k} \in \mathbb{C}^{M \times M}$ is given by 
\begin{equation}
    \label{eq.CE.digital_hk_MMSE_Adk}
    \mathbf{A}_{d,k} = \sqrt{p_k}\mathbf{R}_{k}(p_kT_u\mathbf{R}_{k}+\sigma^2\mathbf{I}_{M})^{-1}.
\end{equation}
The resulting \ac{MSE} is 
\begin{align}
    & \mathbb{E}\{\|\mathbf{h}_{k}-\mathbf{h}_k^{\text{MMSE}}\|^2\}\\
    & = \mathrm{tr}\Bigl(\mathbf{R}_{k} - p_kT_u\mathbf{R}_{k}\bigl(p_kT_u\mathbf{R}_{k}+\sigma^2\mathbf{I}_{M}\bigr)^{-1}\mathbf{R}_{k}\Bigr) \label{eq.CE.digital_hk_MMSE_MSE}.
\end{align}
Finally, by collecting \eqref{eq.CE.digital_hk_LS} and \eqref{eq.CE.digital_hk_MMSE} over all $K$ users, we obtain the estimated channel matrices $\mathbf{H}^{\text{LS}}$ and $\mathbf{H}^{\text{MMSE}}$.

We now compare the \ac{MiLAC}-aided schemes with the conventional digital scheme from two perspectives: \ac{MSE} and computational complexity. 
In terms of \ac{MSE}, for the \ac{LS} estimator, \eqref{eq.CE.hvk_LS_MSE}, \eqref{eq.CE.hvktilde_LS_MSE}, and \eqref{eq.CE.digital_hk_LS_MSE} show that the resulting \ac{MSE} is proportional to the dimension of the channel vector being estimated. 
Specifically, the schemes in Sections~\ref{Sec.3.2} and \ref{Sec.3.3} estimate the $r_g$-dimensional virtual channel $\mathbf{h}_{v,k}$ and the $\tilde{r}$-dimensional global virtual channel $\tilde{\mathbf{h}}_{v,k}$, respectively, while the digital scheme estimates the $M$-dimensional channel $\mathbf{h}_k$. 
Since $r_g \leq \tilde{r} \leq M$, the corresponding \ac{LS} \ac{MSE}s follow the same ordering. 
This \ac{MSE} reduction over the digital scheme comes from the analog projection performed by the \ac{MiLAC}, which removes noise components outside the subspace of interest \cite{Holographic_MIMO}.
Moreover, as $G$ rises from $1$ to $K$, the dimension $\tilde{r}=\mathrm{rank}(\mathbf{U})$ grows from $r_g$ toward $M$. Accordingly, the \ac{LS} \ac{MSE} of estimating $\tilde{\mathbf{h}}_{v,k}$ varies from that of estimating $\mathbf{h}_{v,k}$ to that of estimating $\mathbf{h}_k$. In contrast, for the \ac{MMSE} estimator, the \ac{MiLAC}-aided schemes achieve the same \ac{MSE} as the digital scheme, as proven in Appendix~\ref{Ap.MMSE_MSE_equivalence}. 

In terms of computational complexity, we first derive the complexity of the conventional digital scheme. Recall that $T_u = K$. For the \ac{LS} estimator, the complexity is dominated by the matrix-vector product $\mathbf{Y}\bar{\mathbf{x}}_{k}^*$ for all $K$ users, requiring $8K^2M$ real operations, while the subsequent entry-wise scaling of $\mathbf{z}_{d,k}$ is negligible. For the \ac{MMSE} estimator, the complexity becomes $8(K^2M+KM^2)$ real operations, where the latter term arises from the matrix-vector product $\mathbf{A}_{d,k}\mathbf{z}_{d,k}$ for all $K$ users, with $\mathbf{A}_{d,k}$ precomputed offline given $\mathbf{R}_k$.

\begin{table}[!t]
\caption{A Summary of Computational Complexity.}
\label{tab.1}
\centering
\renewcommand{\arraystretch}{1.25}
\begin{tabular}{|c|c|c|}
\hline
\multirow{2}{*}{Scheme} & \multicolumn{2}{c|}{Computational Complexity} \\
\cline{2-3}
& LS & MMSE \\
\hline
Virtual CE (small $G$)
& $8rK^2/G$ 
& $8rK^2/G$ \\
\hline
Global Virtual CE (large $G$)
& $8\tilde{r}K^2$ 
& $8(\tilde{r}K^2+\tilde{r}^2K)$ \\
\hline
Conventional Digital CE 
& $8K^2M$ 
& $8(K^2M+KM^2)$ \\
\hline
\end{tabular}
\begin{flushleft}
\footnotesize CE: channel estimation.
\end{flushleft}
\end{table}

The computational complexity of all schemes using either the \ac{LS} or \ac{MMSE} estimator is summarized in Table.~\ref{tab.1}, from which several observations can be made. First, when $G$ is small, i.e., $r=\mathrm{rank}(\mathbf{U})<M$, the virtual channel estimation scheme achieves significantly lower computational complexity than the digital one. The gain is particularly pronounced for the \ac{MMSE} estimator due to the additional complexity term that scales quadratically with $M$ in the digital scheme. Second, when $G$ is large, i.e., $\tilde r=\mathrm{rank}(\mathbf{U})\leq \min\{M,r\}$, the global virtual channel estimation scheme still achieves lower computational complexity than the digital one, although the gain becomes smaller than in the case with small $G$. 

Apart from these advantages, the \ac{MiLAC}-aided schemes require only a limited number of \ac{RF} chains, i.e., $L=K$, which is much smaller than the full \ac{RF} chains required by the digital scheme, i.e, $L=M$. This can lead to significant savings in hardware cost and power consumption. However, these benefits do not come for free, since they introduce extra training overhead. Compared with the digital scheme, which uses $T_u=K$, the proposed \ac{MiLAC}-aided schemes require a training overhead of $T_u=K\lceil r/K\rceil$ for small $G$ and $T_u=K\lceil \tilde{r}/K\rceil$ for large $G$. Although the extra training overhead does not affect the accuracy of uplink channel estimation, it reduces the number of symbol durations available for downlink data transmission, whose impact is presented in Section~\ref{Sec.4}.

Overall, the proposed \ac{MiLAC}-aided schemes can achieve several advantages over the conventional digital scheme in terms of \ac{MSE}, computational complexity, and \ac{RF}-chain requirement. The former two are especially pronounced when $G$ is small, and may degrade as $G$ approaches $K$. More detailed numerical results are provided in Section~\ref{Sec.5}.

\section{MiLAC-Aided Beamforming}
\label{Sec.4} 
In this section, we develop \ac{MiLAC}-aided \ac{R-ZFBF} schemes based on the virtual channel $\mathbf{H}_v$ and the global virtual channel $\tilde{\mathbf{H}}_v$ obtained in Sections~\ref{Sec.3.2} and \ref{Sec.3.3}, respectively. Specifically, we introduce a cascade \ac{MiLAC} architecture formed by connecting two \ac{MiLAC}s in series, such that the \ac{R-ZFBF} precoder can be realized directly from the low-dimensional channel estimates with low computational complexity.

\subsection{Motivation for Cascade MiLAC}
\label{Sec.4.1}
Under \ac{TDD} operation, the uplink and downlink channels are reciprocal, so the downlink channel is given by $\mathbf{H}^T$. Accordingly, the \ac{R-ZFBF} precoder is expressed as
\begin{equation}
    \label{eq.BF.W_H}
    \mathbf{W} = \mathbf{H}^*(\mathbf{H}^T\mathbf{H}^*+\lambda\mathbf{I}_K)^{-1},
\end{equation}
where $\lambda=\sigma^2K/P_T$ \cite{MiLAC_Part2}. Here, we drop the time index $t$ for notational simplicity, since the precoder is designed once based on the acquired \ac{CSI} and remains fixed over $T_d$ symbol durations within each coherence block. Recalling \eqref{eq.CE.H_intermsof_Hv} and \eqref{eq.CE.H_intermsof_Hvtilde}, the downlink channel can be equivalently expressed as
\begin{equation}
    \label{eq.BF.HT_intermsof_HvT_tildeHvT}
    \mathbf{H}^T = \mathbf{H}_v^T\mathbf{U}^T = \tilde{\mathbf{H}}_v^T\tilde{\mathbf{U}}^T, 
\end{equation}
where $\mathbf{H}_v^T \in \mathbb{C}^{K\times r}$ and $\tilde{\mathbf{H}}_v^T \in \mathbb{C}^{K\times \tilde r}$ denote the downlink virtual channel and downlink global virtual channel, respectively. Substituting \eqref{eq.BF.HT_intermsof_HvT_tildeHvT} into \eqref{eq.BF.W_H} yields
\begin{equation}
    \label{eq.BF.W_Hv}
    \mathbf{W} = \mathbf{U}^*\mathbf{H}_v^*(\mathbf{H}_{v}^T\mathbf{U}^T\mathbf{U}^*\mathbf{H}_v^*+\lambda\mathbf{I}_K)^{-1},
\end{equation}
and 
\begin{equation}
    \label{eq.BF.W_Hvtilde}
    \mathbf{W} = \tilde{\mathbf{U}}^*\tilde{\mathbf{H}}_v^*(\tilde{\mathbf{H}}_{v}^T\tilde{\mathbf{H}}_v^*+\lambda\mathbf{I}_K)^{-1},   
\end{equation}
where \eqref{eq.BF.W_Hvtilde} follows from $\tilde{\mathbf{U}}^H\tilde{\mathbf{U}}=\mathbf{I}_{\tilde r}$. These expressions show that the \ac{R-ZFBF} precoder can be constructed directly from either $\mathbf{H}_v$ or $\tilde{\mathbf{H}}_v$, without explicitly reconstructing $\mathbf{H}$.

Realizing \eqref{eq.BF.W_Hv} or \eqref{eq.BF.W_Hvtilde} using a single \ac{MiLAC} is computationally inefficient. To see this, we recall the most general form of the precoder that a single \ac{MiLAC} can implement while retaining low computational complexity, as established in \cite{MiLAC_Part1,MiLAC_Part2}. Define the auxiliary matrix $\mathbf{P}\in\mathbb{C}^{(L+M)\times(L+M)}$ associated with the MiLAC admittance matrix $\mathbf{Y}_W$ in \eqref{eq.SM.W_downlink} as
\begin{equation}
    \label{eq.BF.P_single_milac}
    \mathbf{P}=\frac{\mathbf{Y}_{W}}{Y_0}+\mathbf{I}_{L+M}
    =
    \begin{bmatrix}
         \mathbf{P}_{11} & \mathbf{P}_{12} \\
         \mathbf{P}_{21} & \mathbf{P}_{22}
    \end{bmatrix},
\end{equation}
where $\mathbf{P}_{11} \in \mathbb{C}^{L\times L}$, $\mathbf{P}_{12} \in \mathbb{C}^{L\times M}$, $\mathbf{P}_{21} \in \mathbb{C}^{M\times L}$, and $\mathbf{P}_{22} \in \mathbb{C}^{M\times M}$. Assuming that $\mathbf{P}$ and $\mathbf{P}_{22}$ are invertible, then the precoder implemented by a single \ac{MiLAC} must take the form 
\begin{equation}
    \label{eq.BF.W_single_milac}
    \mathbf{W}
    =\mathbf{P}_{22}^{-1}\mathbf{P}_{21} \bigl( \mathbf{P}_{12}\mathbf{P}_{22}^{-1}\mathbf{P}_{21} -\mathbf{P}_{11} \bigr)^{-1}.
\end{equation}
Once the submatrices of $\mathbf{P}$ are known, \eqref{eq.BF.W_single_milac} can be realized with only $6LM$ real operations for computing the admittance components of the single MiLAC \cite{MiLAC_Part2}. A notable example is \eqref{eq.BF.W_H}, obtained by setting $\mathbf{P}_{11}=\pm\lambda\mathbf{I}_{L}$, $\mathbf{P}_{12}=\mathbf{H}^T$, $\mathbf{P}_{21}=\mathbf{H}^*$, and $\mathbf{P}_{22}=\mp\mathbf{I}_{M}$. However, this is no longer possible when only $\mathbf{H}_v$ or $\tilde{\mathbf{H}}_v$ is available. In this case, realizing \eqref{eq.BF.W_Hv} or \eqref{eq.BF.W_Hvtilde} using a single \ac{MiLAC} requires computing $\mathbf{H}_v^T\mathbf{U}^T$ or $\tilde{\mathbf{H}}_v^T\tilde{\mathbf{U}}^T$, which is equivalent to reconstructing $\mathbf{H}^T$, requiring $8M\sum_{g \in \mathcal{G}} r_gK_g$ and $8\tilde{r}KM$ real operations, respectively. This severely compromises the low-complexity benefit using \ac{MiLAC}, motivating the introduction of the cascade \ac{MiLAC}.

\subsection{Model and Analysis of Cascade \ac{MiLAC}}
\label{Sec.4.2}

\begin{figure}[!t]
\centering
\includegraphics[width=0.45\columnwidth]{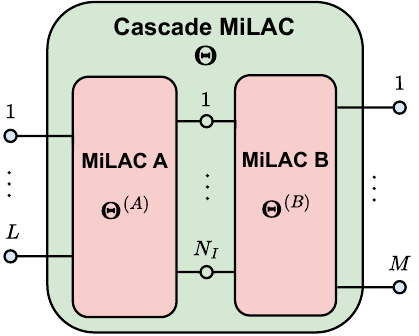}
\caption{Representation of an $(L+M)$-port cascade MiLAC}
\label{fig.3}
\end{figure}

We define a cascade \ac{MiLAC} as a \ac{MiLAC} formed by interconnecting several \ac{MiLAC}s in series through internal ports, such that its scattering matrix is determined by those of its constituents 
\footnote{We use scattering matrices since they provide a direct mapping between the constituent MiLACs and the precoder realized by the cascade MiLAC, as shown in \eqref{eq.BF_W_intermsof_Theta_subblocks}. An equivalent derivation in terms of admittance matrices exists but it yields more cumbersome expressions and obscures this mapping.}.
Compared with a single \ac{MiLAC}, it provides a more general architecture with greater flexibility for low-complexity analog computation. In this work, we focus on a cascade of two \ac{MiLAC}s, which suffices to realize \eqref{eq.BF.W_Hv} and \eqref{eq.BF.W_Hvtilde}. Extension to more general cascade is left for future work.

As illustrated in Fig.~\ref{fig.3}, we consider a cascade of an $(L+N_{I})$-port \ac{MiLAC} and an $(N_I+M)$-port \ac{MiLAC}, with $L\leq N_I\leq M$. Their scattering matrices, $\mathbf{\Theta}^{(A)} \in \mathbb{C}^{(L+N_{I}) \times (L+N_{I})}$ and $\mathbf{\Theta}^{(B)} \in \mathbb{C}^{(N_I+M) \times (N_I+M)}$, are partitioned as
\begin{align}
    \mathbf{\Theta}^{(A)} &= 
    \left[\begin{array}{cc}
         \mathbf{\Theta}^{(A)}_{11} & \mathbf{\Theta}^{(A)}_{12} \\
         \mathbf{\Theta}^{(A)}_{21} & \mathbf{\Theta}^{(A)}_{22}
    \end{array}\right], \label{eq.BF_Theta_A}\\
    \mathbf{\Theta}^{(B)} &= 
    \left[\begin{array}{cc}
         \mathbf{\Theta}^{(B)}_{11} & \mathbf{\Theta}^{(B)}_{12} \\
         \mathbf{\Theta}^{(B)}_{21} & \mathbf{\Theta}^{(B)}_{22}
    \end{array}\right], \label{eq.BF_Theta_B}
\end{align}
where $\mathbf{\Theta}^{(A)}_{11} \in \mathbb{C}^{L \times L}$, $\mathbf{\Theta}^{(A)}_{12} \in \mathbb{C}^{L \times N_I}$, $\mathbf{\Theta}^{(A)}_{21} \in \mathbb{C}^{N_I \times L}$, $\mathbf{\Theta}^{(A)}_{22} \in \mathbb{C}^{N_I \times N_I}$, $\mathbf{\Theta}^{(B)}_{11} \in \mathbb{C}^{N_I \times N_I}$, $\mathbf{\Theta}^{(B)}_{12} \in \mathbb{C}^{N_I \times M}$, $\mathbf{\Theta}^{(B)}_{21} \in \mathbb{C}^{M \times N_I}$, and $\mathbf{\Theta}^{(B)}_{22} \in \mathbb{C}^{M \times M}$.
The last $N_I$ ports of the $(L+N_{I})$-port \ac{MiLAC} are connected to the first $N_I$ ports of the $(N_I+M)$-port \ac{MiLAC}. Thus, the whole cascade can be regarded as an $(L+M)$-port \ac{MiLAC} \cite[Proposition~1]{SIM_Physic_Model}, with scattering matrix $\mathbf{\Theta} \in \mathbb{C}^{(L+M) \times (L+M)}$ given by
\begin{equation}
    \label{eq.BF_Theta}
    \mathbf{\Theta} = 
    \left[\begin{array}{cc}
         \mathbf{\Theta}_{11} & \mathbf{\Theta}_{12} \\
         \mathbf{\Theta}_{21} & \mathbf{\Theta}_{22}
    \end{array}\right],
\end{equation}
where $\mathbf{\Theta}_{11}\in\mathbb{C}^{L\times L}$, $\mathbf{\Theta}_{12}\in\mathbb{C}^{L\times M}$, $\mathbf{\Theta}_{21}\in\mathbb{C}^{M\times L}$, and $\mathbf{\Theta}_{22}\in\mathbb{C}^{M\times M}$ satisfy
\begin{align}
    \mathbf{\Theta}_{11} &= \mathbf{\Theta}^{(A)}_{11}+\mathbf{\Theta}^{(A)}_{12} (\mathbf{I}_{N_I}-\mathbf{\Theta}^{(B)}_{11}\mathbf{\Theta}^{(A)}_{22})^{-1}\mathbf{\Theta}^{(B)}_{11}\mathbf{\Theta}^{(A)}_{21}, \label{eq.BF_Theta_11}\\
    \mathbf{\Theta}_{12} &= \mathbf{\Theta}^{(A)}_{12}(\mathbf{I}_{N_I}-\mathbf{\Theta}^{(B)}_{11}\mathbf{\Theta}^{(A)}_{22})^{-1}\mathbf{\Theta}^{(B)}_{12}, \label{eq.BF_Theta_12}\\  
    \mathbf{\Theta}_{21} &= \mathbf{\Theta}^{(B)}_{21}(\mathbf{I}_{N_I}-\mathbf{\Theta}^{(A)}_{22}\mathbf{\Theta}^{(B)}_{11})^{-1}\mathbf{\Theta}^{(A)}_{21}, \label{eq.BF_Theta_21}\\     
    \mathbf{\Theta}_{22} &= \mathbf{\Theta}^{(B)}_{22}+\mathbf{\Theta}^{(B)}_{21}(\mathbf{I}_{N_I}-\mathbf{\Theta}^{(A)}_{22}\mathbf{\Theta}^{(B)}_{11})^{-1}\mathbf{\Theta}^{(A)}_{22}\mathbf{\Theta}^{(B)}_{12}.\label{eq.BF_Theta_22}
\end{align}

To characterize the precoder realized by the cascade \ac{MiLAC}, we relate its scattering matrix $\mathbf{\Theta}$ to the corresponding admittance matrix $\mathbf{Y}_W$. Refer to \cite[Chapter~4]{Microwave_Engineering_Book}, we have
\begin{equation}
    \label{eq.BF_Theta_Y_Basic}
    \mathbf{\Theta} = (Y_0\mathbf{I}_{L+M}+\mathbf{Y}_W)^{-1}(Y_0\mathbf{I}_{L+M}-\mathbf{Y}_W),
\end{equation}
which implies
\begin{equation}
    \label{eq.BF_Theta_Y_Mid}
    (\frac{\mathbf{Y}_W}{Y_0}+\mathbf{I}_{L+M})^{-1} = \frac{1}{2}(\mathbf{\Theta}+\mathbf{I}_{L+M}),
\end{equation}
as established in \cite[(27)]{MiLAC_Capcity}. Thus, we can express the precoder $\mathbf{W}$ in \eqref{eq.SM.W_downlink} in terms of $\mathbf{\Theta}$ as  
\begin{equation}
    \label{eq.BF_W_intermsof_Theta}
    \mathbf{W} = \frac{1}{2}[\mathbf{\Theta}]_{L+\mathcal{M}, \mathcal{L}}\;.
\end{equation}
Since $[\mathbf{\Theta}]_{L+\mathcal{M},\mathcal{L}}=\mathbf{\Theta}_{21}$, it follows from \eqref{eq.BF_Theta_21} that
\begin{equation}
    \label{eq.BF_W_intermsof_Theta_subblocks}
    \mathbf{W} = \frac{1}{2} \mathbf{\Theta}^{(B)}_{21}(\mathbf{I}_{N_I}-\mathbf{\Theta}^{(A)}_{22}\mathbf{\Theta}^{(B)}_{11})^{-1}\mathbf{\Theta}^{(A)}_{21}.
\end{equation}
To ensure only forward transmission in the cascade MiLAC, it is sufficient to impose $\mathbf{\Theta}^{(B)}_{11} = \mathbf{0}_{N_I}$, so that \eqref{eq.BF_W_intermsof_Theta_subblocks} reduces to 
\begin{equation}
    \label{eq.BF_W_intermsof_Theta_Subblocks_Final}
    \mathbf{W} = \frac{1}{2} \mathbf{\Theta}^{(B)}_{21}\mathbf{\Theta}^{(A)}_{21}.
\end{equation}
In the following, we show how to construct the two \ac{MiLAC}s in the form of \eqref{eq.BF.P_single_milac} such that the precoder realized by the resulting cascade \ac{MiLAC} in \eqref{eq.BF_W_intermsof_Theta_Subblocks_Final} can be equivalent to \eqref{eq.BF.W_Hv} or \eqref{eq.BF.W_Hvtilde}, while retaining low computational complexity. 

\subsection{\ac{R-ZFBF} Based on Virtual Channel}
\label{Sec.4.3}

Based on the virtual channel $\mathbf{H}_v$ estimated in Section~\ref{Sec.3.2}, we now design a cascade \ac{MiLAC} to realize \ac{R-ZFBF} for the case with a small number of groups $G$, where $r=\mathrm{rank}(\mathbf{U})<M$. To this end, recalling that $L=K$, we first rewrite \eqref{eq.BF.W_Hv} in a factorized form consistent with \eqref{eq.BF_W_intermsof_Theta_Subblocks_Final} as
\begin{equation}
    \label{eq.BF_W_Extended_Hv}
    \mathbf{W} = \frac{1}{2}\underbrace{(\mathbf{U}^T)^\dagger}_{\mathbf{B} \in \mathbb{C}^{M \times r}}  \underbrace{2\mathbf{U}^T\mathbf{U}^*\mathbf{H}_v^*(\mathbf{H}_{v}^T\mathbf{U}^T\mathbf{U}^*\mathbf{H}_v^*+\lambda\mathbf{I}_K)^{-1}}_{\mathbf{A}\in \mathbb{C}^{r \times L}},
\end{equation}
where $(\mathbf{U}^T)^\dagger = \mathbf{U}^*(\mathbf{U}^T\mathbf{U}^*)^{-1}$. To realize \eqref{eq.BF_W_Extended_Hv}, the scattering matrices of the $(L+N_{I})$-port \ac{MiLAC} and the $(N_I+M)$-port \ac{MiLAC} are required to take the forms
\begin{align}
    \mathbf{\Theta}^{(A)} &= 
    \left[\begin{array}{cc}
         \mathbf{\Theta}^{(A)}_{11} & \mathbf{\Theta}^{(A)}_{12} \\
         \mathbf{A} & \mathbf{\Theta}^{(A)}_{22}
    \end{array}\right], \label{eq.BF_Theta_A_Hv}\\
    \mathbf{\Theta}^{(B)} &= 
    \left[\begin{array}{cc}
         \mathbf{0}_{N_I} & \mathbf{\Theta}^{(B)}_{12} \\
         \mathbf{B} & \mathbf{\Theta}^{(B)}_{22}
    \end{array}\right], \label{eq.BF_Theta_B_Hv}
\end{align}
where $N_I = r$. By further exploiting \eqref{eq.BF_Theta_Y_Mid}, \eqref{eq.BF_Theta_A_Hv} and \eqref{eq.BF_Theta_B_Hv} can be equivalently expressed as
\begin{align}
    (\mathbf{P}^{(A)})^{-1} &= 
    \frac{1}{2}\left[\begin{array}{cc}
         \mathbf{\Theta}^{(A)}_{11}+\mathbf{I}_{L} & \mathbf{\Theta}^{(A)}_{12} \label{eq.BF_P_inverse_A_Hv}\\
         \mathbf{A} & \mathbf{\Theta}^{(A)}_{22}+\mathbf{I}_{N_I}
    \end{array}\right], \\
    (\mathbf{P}^{(B)})^{-1} &= 
    \frac{1}{2}\left[\begin{array}{cc}
         \mathbf{I}_{N_I} & \mathbf{\Theta}^{(B)}_{12} \\
         \mathbf{B} & \mathbf{\Theta}^{(B)}_{22} +\mathbf{I}_M
    \end{array}\right], \label{eq.BF_P_inverse_B_Hv}
\end{align}
where, following \eqref{eq.BF.P_single_milac}, we introduce
\begin{align}
    \mathbf{P}^{(A)} &= \frac{\mathbf{Y}_W^{(A)}}{Y_0}+\mathbf{I}_{L+N_I} =
    \begin{bmatrix}
         \mathbf{P}_{11}^{(A)} & \mathbf{P}_{12}^{(A)} \\
         \mathbf{P}_{21}^{(A)} & \mathbf{P}_{22}^{(A)}
    \end{bmatrix}, \\
    \mathbf{P}^{(B)} &= \frac{\mathbf{Y}_W^{(B)}}{Y_0}+\mathbf{I}_{N_I+M} =
    \begin{bmatrix}
         \mathbf{P}_{11}^{(B)} & \mathbf{P}_{12}^{(B)} \\
         \mathbf{P}_{21}^{(B)} & \mathbf{P}_{22}^{(B)}
    \end{bmatrix},
\end{align}
with $\mathbf{Y}_W^{(A)}$ and $\mathbf{Y}_W^{(B)}$ denoting the admittance matrices of the $(L+N_I)$-port and $(N_I+M)$-port \ac{MiLAC}s, respectively.

To verify that \eqref{eq.BF_Theta_A_Hv} and \eqref{eq.BF_Theta_B_Hv} are realizable, it suffices to prove the existence of $\mathbf{P}^{(A)}$ and $\mathbf{P}^{(B)}$ satisfying \eqref{eq.BF_P_inverse_A_Hv} and \eqref{eq.BF_P_inverse_B_Hv}. Since $\mathbf{Y}_W^{(A)}$ and $\mathbf{Y}_W^{(B)}$ are assumed fully reconfigurable, the blocks in \eqref{eq.BF_P_inverse_A_Hv} and \eqref{eq.BF_P_inverse_B_Hv}, apart from $\mathbf{A}$, $\mathbf{I}_{N_I}$, and $\mathbf{B}$, can take arbitrary complex values provided that $\mathbf{P}^{(A)}$ and $\mathbf{P}^{(B)}$ remain invertible. Using the $2\times2$ block matrix inversion theorem \cite[Theorem~2.1]{2by2_matrix_inversion}, the problem reduces to a feasibility check. For the $(L+N_{I})$-port \ac{MiLAC}, we need to solve
\begin{align}
    \text{find}\;\;  &\mathbf{P}_{11}^{(A)},\mathbf{P}_{12}^{(A)},\mathbf{P}_{21}^{(A)},\mathbf{P}_{22}^{(A)}\\
    \mathsf{\text{s.t.}}\;\;\;
    & \mathbf{P}_{22}^{(A)} \;\text{is invertible}, \label{eq.P_A_22_invertible}\\
    & \bigl(\mathbf{P}_{12}^{(A)}(\mathbf{P}_{22}^{(A)})^{-1}\mathbf{P}_{21}^{(A)} - \mathbf{P}_{11}^{(A)}\bigr) \; \text{is invertible}, \label{eq.P_A_22_Schur_Complement_invertible}\\
    & (\mathbf{P}_{22}^{(A)})^{-1}\mathbf{P}_{21}^{(A)}\bigl(\mathbf{P}_{12}^{(A)}(\mathbf{P}_{22}^{(A)})^{-1}\mathbf{P}_{21}^{(A)} - \mathbf{P}_{11}^{(A)}\bigr)^{-1} = \frac{1}{2}\mathbf{A},
\end{align}
where \eqref{eq.P_A_22_invertible} and \eqref{eq.P_A_22_Schur_Complement_invertible} are sufficient conditions for $\mathbf{P}^{(A)}$ to be invertible. One valid solution is given by   
\begin{equation}
    \label{eq.BF.PA_Hv}
    \mathbf{P}^{(A)} = 
    \left[\begin{array}{cc}
         \pm\lambda\mathbf{I}_K & \mathbf{H}_v^T \\
          \mathbf{H}_v^*& \mp (\mathbf{U}^T\mathbf{U}^*)^{-1}
    \end{array}\right].
\end{equation}
Similarly, for the $(N_I+M)$-port \ac{MiLAC}, the feasibility check problem is formulated as
\begin{align}
    \text{find}\;\;  &\mathbf{P}_{11}^{(B)},\mathbf{P}_{12}^{(B)},\mathbf{P}_{21}^{(B)},\mathbf{P}_{22}^{(B)}\\
    \mathsf{\text{s.t.}}\;\;\;
    & \mathbf{P}_{22}^{(B)} \;\text{is invertible},\\
    & \bigl(\mathbf{P}_{12}^{(B)}(\mathbf{P}_{22}^{(B)})^{-1}\mathbf{P}_{21}^{(B)} - \mathbf{P}_{11}^{(B)}\bigr) \; \text{is invertible},\\
    & (-\mathbf{P}_{12}^{(B)}(\mathbf{P}_{22}^{(B)})^{-1}\mathbf{P}_{21}^{(B)} + \mathbf{P}_{11}^{(B)}\bigr)^{-1} = \frac{1}{2}\mathbf{I}_{N_I},\\
    & (\mathbf{P}_{22}^{(B)})^{-1}\mathbf{P}_{21}^{(B)}\bigl(\mathbf{P}_{12}^{(B)}(\mathbf{P}_{22}^{(B)})^{-1}\mathbf{P}_{21}^{(B)} - \mathbf{P}_{11}^{(B)}\bigr)^{-1} = \frac{1}{2}\mathbf{B},
\end{align}
with one valid solution presented as 
\begin{equation}
    \label{eq.BF.PB_Hv}
    \mathbf{P}^{(B)} = 
    \left[\begin{array}{cc}
         2\mathbf{I}_{N_I} & \mathbf{0}_{N_I \times M} \\
         -\mathbf{B} & \mathbf{I}_{M}
    \end{array}\right].
\end{equation}
Overall, with \eqref{eq.BF.PA_Hv} and \eqref{eq.BF.PB_Hv}, the resulting cascade \ac{MiLAC} realizes \eqref{eq.BF_W_Extended_Hv}, which is equivalent to \eqref{eq.BF.W_Hv} for small $G$, thereby establishing \ac{MiLAC}-aided \ac{R-ZFBF} based on $\mathbf{H}_v$.

\begin{remark}
    \label{Remark 3}
    Since $\mathbf{U}$ depends only on the channel statistics, all terms involving it, such as $(\mathbf{U}^T\mathbf{U}^*)^{-1}$ in \eqref{eq.BF.PA_Hv} and $\mathbf{B}$ in \eqref{eq.BF.PB_Hv}, can be precomputed offline. Furthermore, since the data symbol vector is precoded by the cascade \ac{MiLAC} in the analog domain as $\mathbf{x}'_t=\mathbf{W}\mathbf{s}_t$, no digital computation is required per symbol duration. Hence, within each coherence block, the computational complexity of \ac{MiLAC}-aided \ac{R-ZFBF} based on $\mathbf{H}_v$ is dominated by computing the admittance components of the $(L+N_I)$-port \ac{MiLAC} from \eqref{eq.BF.PA_Hv}, requiring $6rK$ real operations, given $N_I=r$ and $L = K$. A comparison with conventional digital \ac{R-ZFBF} is deferred to Remark~\ref{Remark 4}.
\end{remark}

\subsection{\ac{R-ZFBF} Based on Global Virtual Channel}
\label{Sec.4.4}

We now design a cascade \ac{MiLAC} to realize \ac{R-ZFBF} based on the global virtual channel $\tilde{\mathbf{H}}_v$ obtained in Section~\ref{Sec.3.3} for large $G$, where $\tilde{r} = \mathrm{rank}(\mathbf{U}) \leq \min\{M,r\}$. Since the derivation closely parallels that in Section~\ref{Sec.4.3}, we present only the key steps and highlight the differences. First, \eqref{eq.BF.W_Hvtilde} can be rewritten in a factorized form consistent with \eqref{eq.BF_W_intermsof_Theta_Subblocks_Final} as
\begin{equation}
    \label{eq.BF_W_Extended_Hvtilde}
    \mathbf{W} = \frac{1}{2}\underbrace{\tilde{\mathbf{U}}^*}_{\mathbf{B} \in \mathbb{C}^{M \times \tilde{r}}}  \underbrace{2\tilde{\mathbf{H}}_v^*(\tilde{\mathbf{H}}_v^T\tilde{\mathbf{H}}_v^*+\lambda\mathbf{I}_K)^{-1}}_{\mathbf{A}\in \mathbb{C}^{\tilde{r} \times L}}.
\end{equation}
With $\mathbf{A}$ and $\mathbf{B}$ defined above, the same design procedure as in Section~\ref{Sec.4.3} can be applied here, except that $N_I=\tilde r$. As a result, valid choices of $\mathbf{P}^{(A)}$ and $\mathbf{P}^{(B)}$ can be obtained. For the $(L+N_I)$-port single \ac{MiLAC}, one valid choice of $\mathbf{P}^{(A)}$ is obtained by simplifying \eqref{eq.BF.PA_Hv} as
\begin{equation}
    \label{eq.BF.PA_Hvtilde}
    \mathbf{P}^{(A)} = 
    \left[\begin{array}{cc}
         \pm\lambda\mathbf{I}_K & \tilde{\mathbf{H}}_v^T \\
          \tilde{\mathbf{H}}_v^*& \mp \mathbf{I}_{N_I}
    \end{array}\right].
\end{equation}
For the $(N_I+M)$-port single \ac{MiLAC}, a valid choice of $\mathbf{P}^{(B)}$ takes the same form as in \eqref{eq.BF.PB_Hv}, except with $\mathbf{B}=\tilde{\mathbf{U}}^*$. Consequently, the resulting cascade \ac{MiLAC} realizes \eqref{eq.BF_W_Extended_Hvtilde}, which is equivalent to \eqref{eq.BF.W_Hvtilde}, thereby establishing \ac{MiLAC}-aided \ac{R-ZFBF} based on $\tilde{\mathbf{H}}_v$ for large $G$.

\begin{remark}
    \label{Remark 4}
    Following the same reasoning as in Section~\ref{Sec.4.3}, the computational complexity of \ac{MiLAC}-aided \ac{R-ZFBF} based on $\tilde{\mathbf{H}}_v$ is dominated by computing the admittance components of the $(L+N_I)$-port \ac{MiLAC} from \eqref{eq.BF.PA_Hvtilde}. Since $N_I = \tilde{r}$ and $L = K$, this requires $6\tilde{r}K$ real operations per coherence block, comparable to the $6rK$ real operations required when based on $\mathbf{H}_v$. 
    In contrast, the complexity of conventional digital \ac{R-ZFBF} based on $\mathbf{H}$ consists of two parts. The first is the computation of the precoder in \eqref{eq.BF.W_H}, which involves matrix-matrix products and a matrix inversion, requiring $(16K^2M+8K^3/3)$ real operations per coherence block. The second is the digital precoding of the data symbol vector, i.e., $\mathbf{x}'_t=\mathbf{W}\mathbf{s}_t$, in each downlink symbol duration, which requires $8KM$ real operations. Therefore, within each coherence block, the complexity of digital \ac{R-ZFBF} is $8\bigl(2K^2M+K^3/3+KM(T-K))\bigr)$ real operations, which is significantly higher than that of the \ac{MiLAC}-aided \ac{R-ZFBF} schemes. 
    Moreover, the cascade MiLAC designed in Section~\ref{Sec.4.3} and \ref{Sec.4.4} can exactly realize \eqref{eq.BF.W_Hv} and \eqref{eq.BF.W_Hvtilde}, respectively, both equivalent to \eqref{eq.BF.W_H}. Therefore, whether based on $\mathbf{H}_v$ for small $G$, or on $\tilde{\mathbf{H}}_v$ for large $G$, the \ac{MiLAC}-aided \ac{R-ZFBF} achieves the same sum rate as digital \ac{R-ZFBF}. The difference only arises when the training overhead is taken into account. Specifically, the effective sum rate of the MiLAC-aided scheme is scaled by $(1-K\lceil{r}/K\rceil/T)$ and $(1-K\lceil \tilde{r}/K\rceil/T)$ when based on $\mathbf{H}_v$ and $\tilde{\mathbf{H}}_v$, respectively, while the digital R-ZFBF is scaled by $(1-K/T)$. Hence, the \ac{MiLAC}-aided \ac{R-ZFBF} schemes may incur a slight loss in effective sum rate due to the extra training overhead. The trade-off between $G$, computational complexity, and sum rate is numerically studied in Section~\ref{Sec.5}. 
\end{remark}

\section{Numerical Results}
\label{Sec.5}
In this section, we present numerical results to evaluate the performance and computational complexity of the proposed MiLAC-aided channel estimation and beamforming schemes.

\subsection{Simulation Setup}
We consider a \ac{MiLAC}-aided \ac{MU-MISO} system with $K = 16$ single-antenna users and a \ac{BS} equipped with $M = 256$ antennas arranged in a \ac{ULA}. The users are partitioned into $G$ groups, with $G \in \{1,2,4,8,16\}$, while all groups are assumed to have equal size, i.e., $K_g = K/G$. Moreover, the coherence block length is set to $T = 1500$ symbol durations, and the noise power is set to $\sigma^2=-80\,\text{dBm}$. 

The channel vector of user $k\in\mathcal{K}_g$ in group $g \in \mathcal{G}$ is modelled as $\mathbf{h}_{k} = \sqrt{\beta_{k}}\mathbf{U}_g\mathbf{\Lambda}_g^{1/2}\mathbf{v}_{k}$, as in \eqref{eq.SM.hvk}. 
The group correlation matrix $\mathbf{R}_g=\mathbf{U}_g\mathbf{\Lambda}_g\mathbf{U}_g^H$ is generated according to the one-ring model in \cite[Section~2.6]{Massive_MIMO_Book}, with $(i,j)$-th entry 
\begin{equation}
    \label{R_g_entry}
    [\mathbf{R}_g]_{i,j}
    =
    \frac{1}{2\sqrt{3}\sigma_{\theta_g}}
    \int_{-\sqrt{3}\sigma_{\theta_g}}^{\sqrt{3}\sigma_{\theta_g}}
    e^{j2\pi d_A (i-j)\sin(\theta_g+\delta_g)}\, d\delta_g, 
\end{equation}
where $\theta_g$ is the nominal angle, $\delta_g$ is the angular deviation uniformly distributed over $[-\sqrt{3}\sigma_{\theta_g},\sqrt{3}\sigma_{\theta_g}]$, $\sigma_{\theta_g}$ is the \ac{ASD}, and $d_A$ is the antenna spacing of the \ac{ULA} measured in wavelengths. In the simulations, we set $\theta_g = -\pi/2 + (2g-1)\pi/(2G)$, $\sigma_{\theta_g} = 8^\circ$, and $d_A = 1/4$ \cite{JSDM_1,Holographic_MIMO}.
For the large-scale fading, the coefficient of user $k$ is modelled as $\beta_k = C_0d_k^{-\alpha}$, where $C_0 = -30\,\text{dBm}$ is the path loss at a reference distance of $1$m, $\alpha=2.8$ is the path loss exponent, and $d_k$ is the distance between user $k$ and the \ac{BS} \cite{SIM_CE}. To generate $d_k$ for users in group $g$, a reference user is first dropped in an annular region centred at the \ac{BS}, with angle $\theta_g$ and distance $d_g$ uniformly distributed over the area between $200$m and $250$m. The remaining $(K_g-1)$ users are then dropped within a small circle centred at the reference user with radius $d_g\tan(\sqrt{3}\sigma_{\theta_g})/10$, one order of magnitude smaller than the corresponding scattering cluster \cite{JSDM_1}.

Given the above setup, in the following simulations, $G \in \{1,2,4\}$ corresponds to the small-$G$ regime, where $r = \mathrm{rank}(\mathbf{U}) <M$. In this regime, we evaluate the virtual channel estimation scheme in Section~\ref{Sec.3.2} and the corresponding \ac{R-ZFBF} scheme based on $\mathbf{H}_v$ in Section~\ref{Sec.4.3}. In contrast, $G \in \{8, 16\}$ corresponds to the large-$G$ regime, where $\tilde{r} = \mathrm{rank}(\mathbf{U})\leq \min\{M,r\}$. Here, we evaluate the global virtual channel estimation scheme in Section~\ref{Sec.3.3} and the corresponding \ac{R-ZFBF} scheme based on $\tilde{\mathbf{H}}_v$ in Section~\ref{Sec.4.4}.

\subsection{MiLAC-Aided Channel Estimation}

\begin{figure}[!t]
\centering
\subfloat[$G = 2$]{\includegraphics[width=0.495\columnwidth]{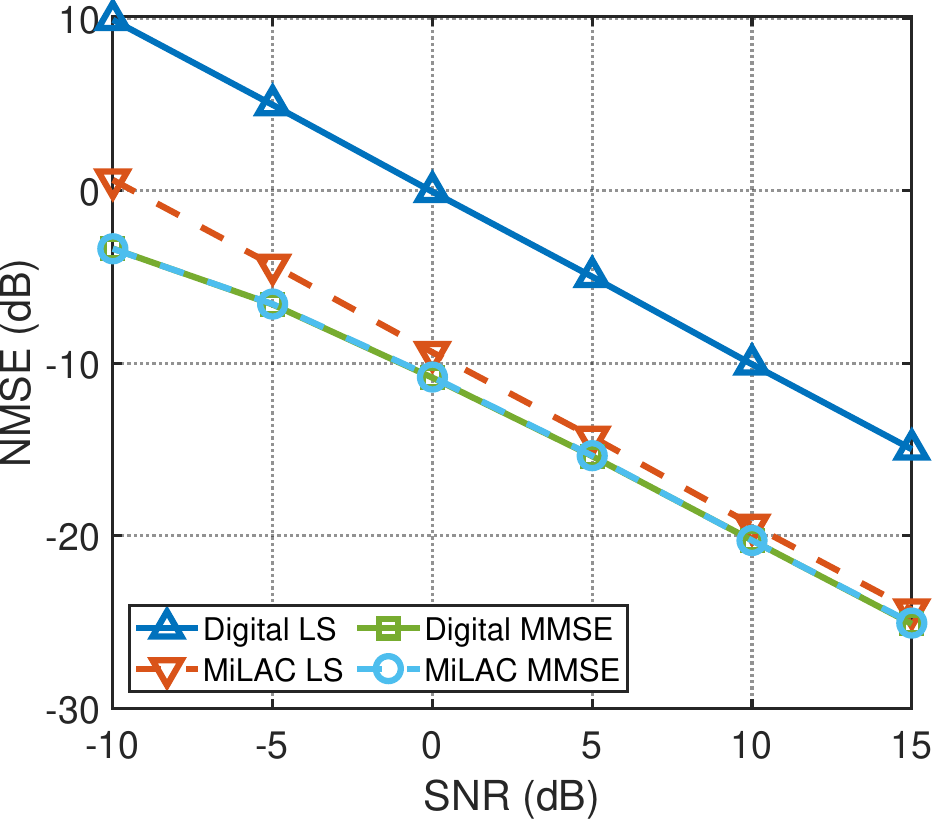}
\label{fig.CE_NMSE.a}}
\subfloat[$G = 8$]{\includegraphics[width=0.495\columnwidth]{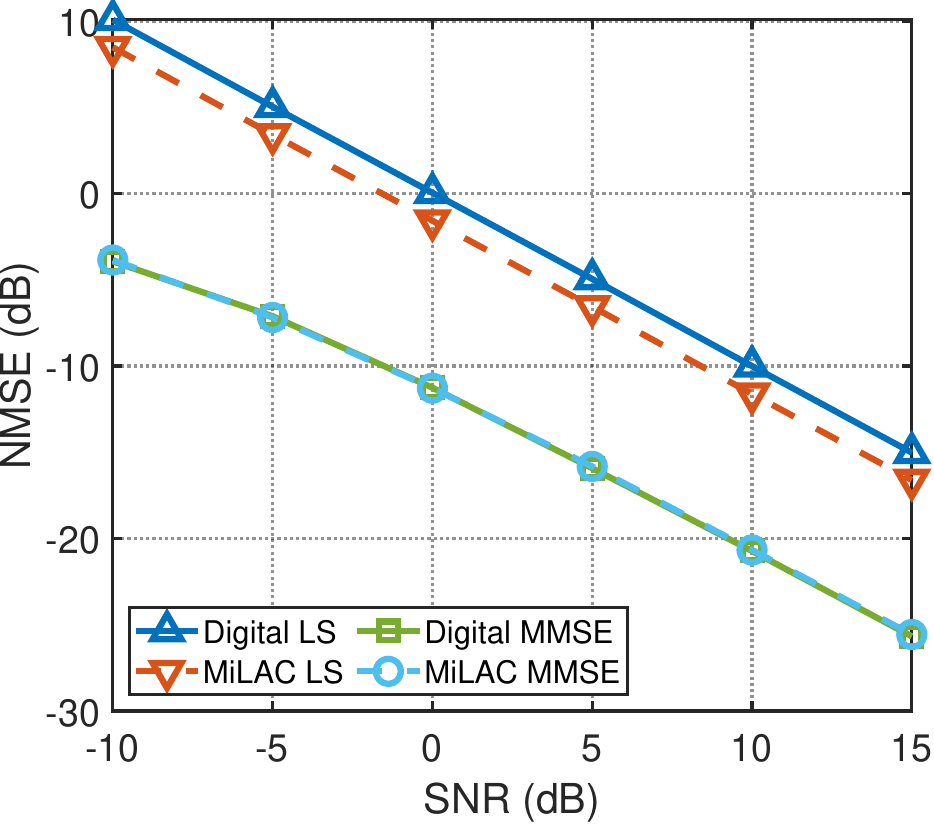}
\label{fig.CE_NMSE.b}}
\caption{Comparison of digital and MiLAC-aided channel estimation: NMSE versus SNR.}
\label{fig.CE_NMSE}
\end{figure}

We first evaluate the performance of the proposed MiLAC-aided channel estimation schemes in Fig.~\ref{fig.CE_NMSE}, by plotting the \ac{NMSE} versus the \ac{SNR}, with digital LS and MMSE included as baselines. Specifically, we consider virtual channel estimation for $G = 2$ and global virtual channel estimation for $G = 8$. The \ac{NMSE} is defined as $\mathrm{NMSE}=\mathbb{E}\{\|\mathbf{H}_\text{rep}-\hat{\mathbf{H}}_\text{rep}\|_F^2\}/\mathbb{E}\{\|\mathbf{H}_\text{rep}\|_F^2\}$, where $\mathbf{H}_{\text{rep}} \in\{\mathbf{H},\mathbf{H}_v,\tilde{\mathbf{H}}_v\}$ is the representation under consideration and $\hat{\mathbf{H}}_{\text{rep}}$ is the corresponding LS or MMSE estimate. The \ac{SNR} is defined as $\mathrm{SNR}_k = p_k\beta_kT_u/\sigma^2$ \cite[(3.13)]{Massive_MIMO_Book}. For fair comparison, $p_k$ is chosen such that $\mathrm{SNR}_k=\mathrm{SNR}$, $\forall k\in\mathcal{K}$. 
From Fig.~\ref{fig.CE_NMSE}, we have the following observations. 
\textit{First}, MiLAC-aided \ac{MMSE} achieves the same performance as digital \ac{MMSE}, as expected in Appendix~\ref{Ap.MMSE_MSE_equivalence}.
\textit{Second}, MiLAC-aided LS outperforms digital LS, consistent with the discussion in Section~\ref{Sec.3.4}.
The gain is pronounced for virtual channel estimation with $G=2$, where MiLAC-aided LS nearly matches the performance of digital MMSE. By contrast, for global virtual channel estimation with $G=8$, the gain shrinks and MiLAC-aided LS approaches the performance of digital LS. This is because the dimension of the global virtual channel increases with $G$, which introduces more noise components into \ac{LS} estimation.

\begin{figure}[!t]
\centering
\subfloat[Small-$G$ Regime]{\includegraphics[width=0.495\columnwidth]{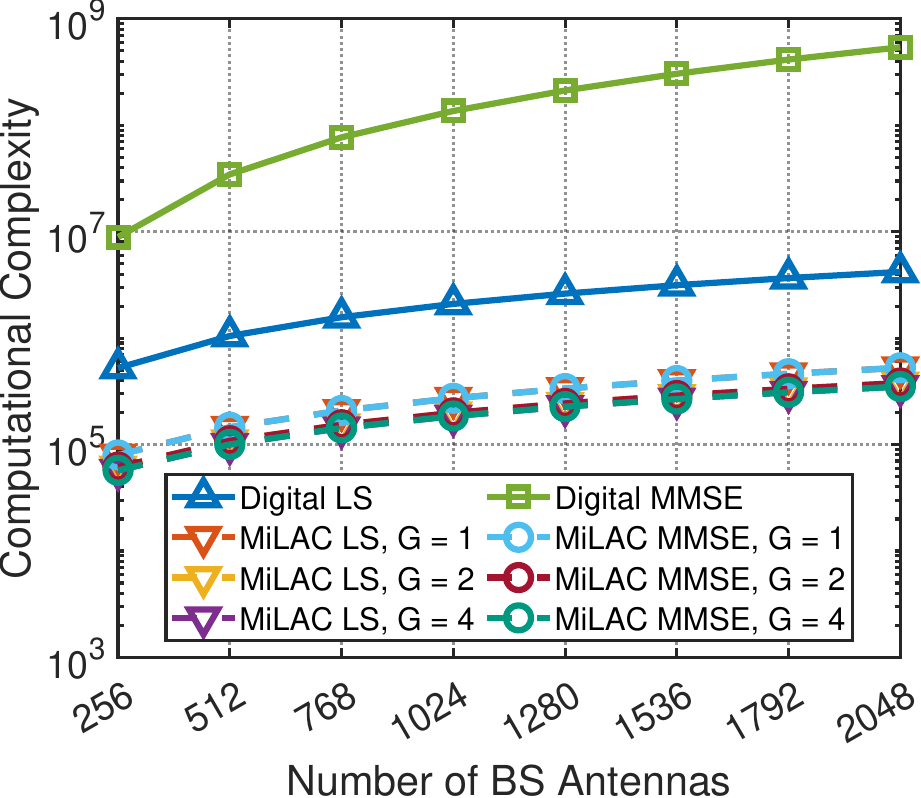}
\label{fig.CE_complexity.a}}
\subfloat[Large-$G$ Regime]{\includegraphics[width=0.495\columnwidth]{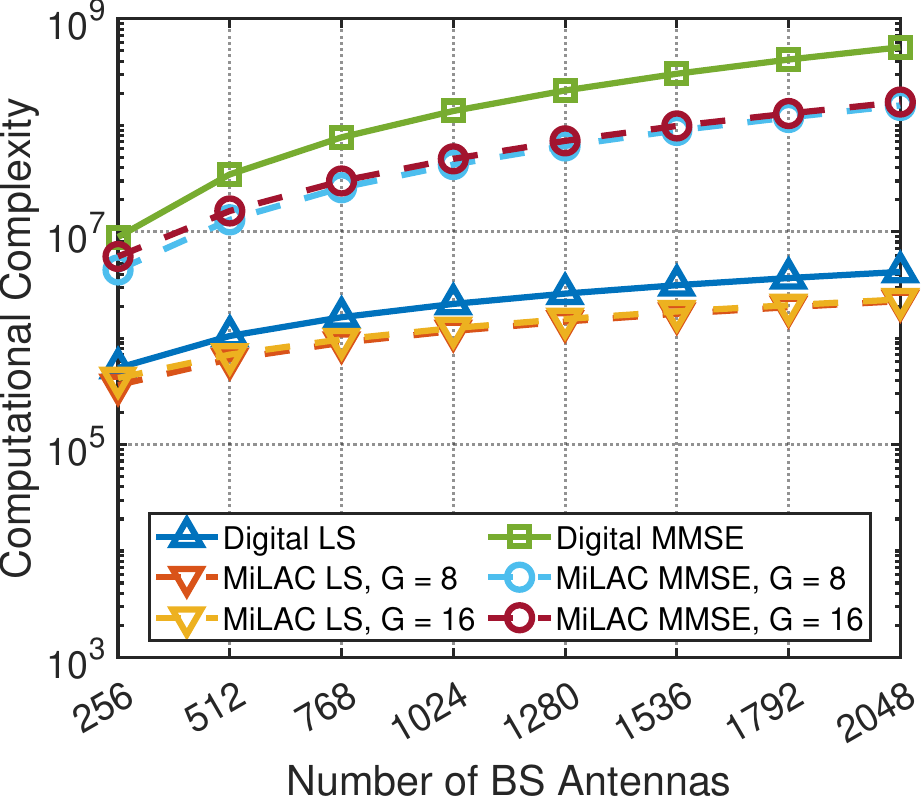}
\label{fig.CE_complexity.b}}
\caption{Comparison of digital and MiLAC-aided channel estimation: computational complexity versus the number of BS antennas.}
\label{fig.CE_complexity}
\end{figure}

We next evaluate the computational complexity of the proposed MiLAC-aided channel estimation schemes (listed in Tab.~\ref{tab.1}) versus the number of \ac{BS} antennas in Fig.~\ref{fig.CE_complexity}, with digital LS and MMSE included for comparison. 
From Fig.~\ref{fig.CE_complexity}, we make the following observations. 
\textit{First}, for virtual channel estimation in the small-$G$ regime, MiLAC-aided LS and MMSE achieve significantly lower computational complexity than their digital counterparts. Particularly, MiLAC-aided MMSE achieves up to a $1540\times$ reduction for $G=4$, corresponding to a saving of $5.41\times10^8$ real operations. Notably, the complexity of MiLAC-aided LS and MMSE decreases when $G$ increases from $1$ to $2$, but changes little from $2$ to $4$. This is because their complexity scales as $8rK^2/G$, where the increase in $G$ is gradually offset by the increase in $r=\sum_{g\in\mathcal{G}}r_g$. 
\textit{Second}, for global virtual channel estimation in the large-$G$ regime, MiLAC-aided LS and MMSE still achieve lower computational complexity than their digital counterparts, although the gains are smaller than those of virtual channel estimation, as expected in Section~\ref{Sec.3.4}. Moreover, the gains vary little with $G$, indicating that $\tilde{r}$ remains unchanged in the large-$G$ regime.

\subsection{MiLAC-Aided Beamforming}

\begin{figure}[!t]
\centering
\subfloat[$G = 2$]{\includegraphics[width=0.495\columnwidth]{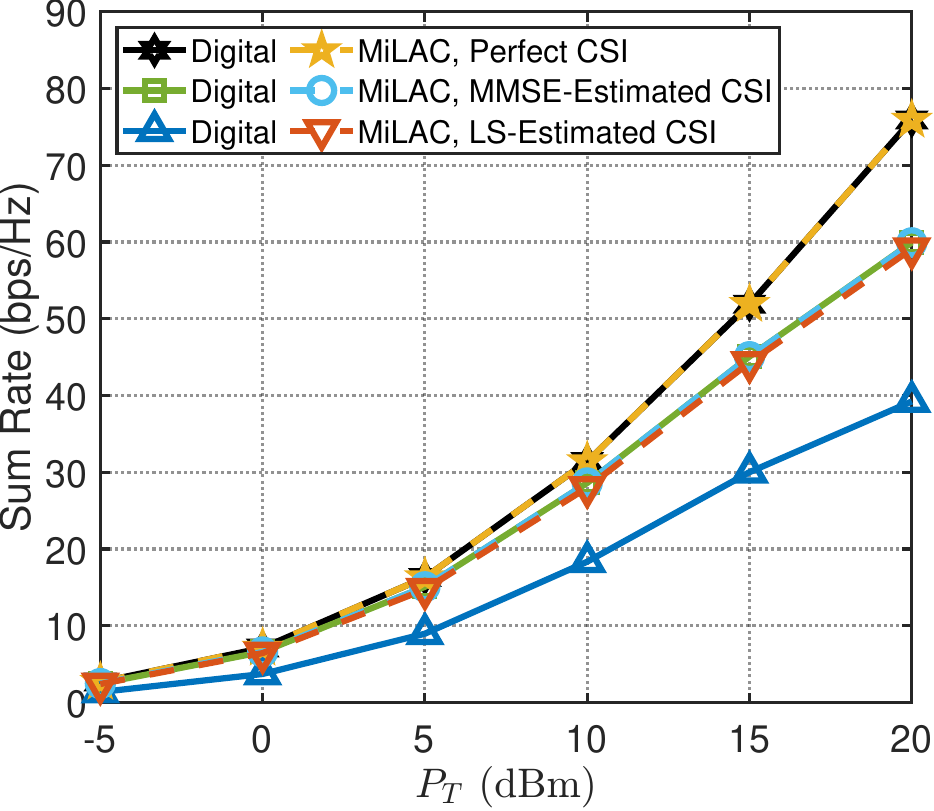}
\label{fig.BF_sum_rate.a}}
\subfloat[$G = 8$]{\includegraphics[width=0.495\columnwidth]{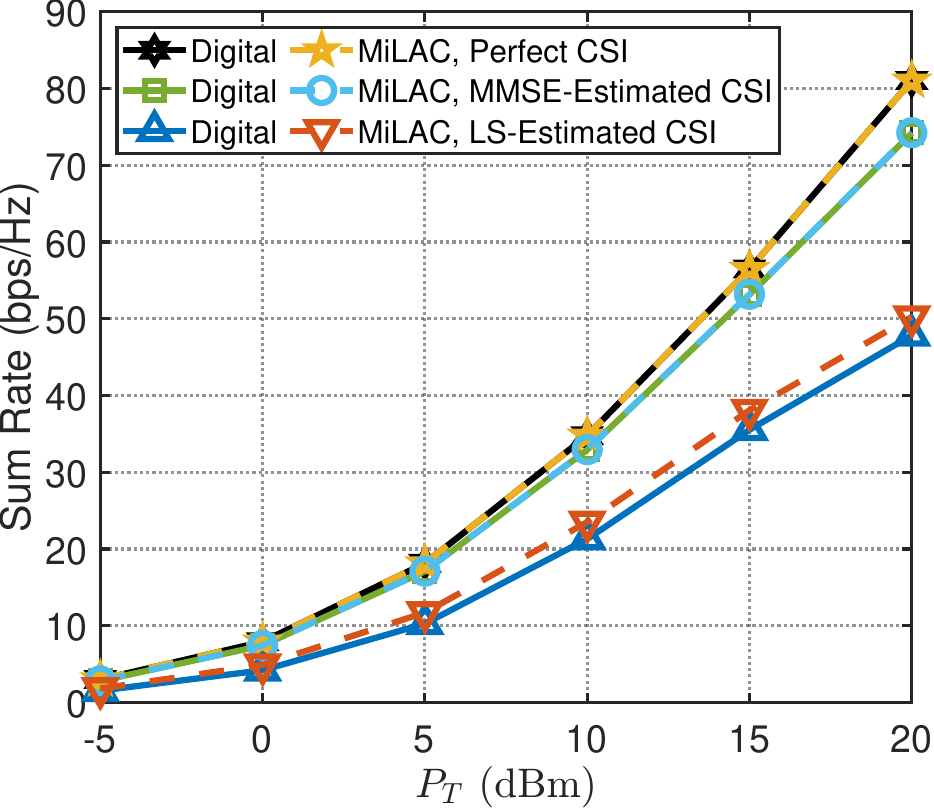}
\label{fig.BF_sum_rate.b}}
\caption{Comparison of digital and MiLAC-aided beamforming: sum rate versus $P_T$.}
\label{fig.BF_sum_rate}
\end{figure}

We start by evaluating the performance of the proposed MiLAC-aided beamforming schemes in Fig.~\ref{fig.BF_sum_rate}, which plots the sum rate versus the average transmit power $P_T$. Specifically, we consider MiLAC-aided \ac{R-ZFBF} based on $\mathbf{H}_v$ for $G = 2$ and $\tilde{\mathbf{H}}_v$ for $G = 8$. In both cases, MiLAC-aided \ac{R-ZFBF} is evaluated using perfect CSI and CSI estimated at $\mathrm{SNR} = 0\,\text{dB}$. The latter is obtained by MiLAC-aided LS and MMSE, corresponding to virtual channel estimation for $G = 2$ or global virtual channel estimation for $G = 8$. As baselines, we include digital \ac{R-ZFBF} based on $\mathbf{H}$, using perfect CSI and CSI estimated at $\mathrm{SNR}=0\,\text{dB}$ by digital LS and MMSE. For a fair comparison, the precoders in \eqref{eq.BF.W_H}, \eqref{eq.BF.W_Hv}, and \eqref{eq.BF.W_Hvtilde} are normalized as $\sqrt{L}\mathbf{W}/\|\mathbf{W}\|_F$, such that the same power is radiated from the \ac{BS} antennas, i.e., $\mathbb{E}[\|\mathbf{x}'_t\|^2] = P_T$. Moreover, training overhead is omitted from the sum rate calculation to  enable a direct comparison between MiLAC-aided and digital \ac{R-ZFBF}.
From Fig.~\ref{fig.BF_sum_rate}, we have the following observations.
\textit{First}, with perfect CSI or MMSE-estimated CSI, MiLAC-aided \ac{R-ZFBF} achieves the same performance as digital \ac{R-ZFBF}. This confirms that the cascade MiLAC can realize \ac{R-ZFBF} based on $\mathbf{H}_v$ or $\tilde{\mathbf{H}}_v$ with the same behaviour as digital \ac{R-ZFBF} based on $\mathbf{H}$, validating our insight in Remark~\ref{Remark 4}.
\textit{Second}, with LS-estimated CSI, MiLAC-aided \ac{R-ZFBF} outperforms digital \ac{R-ZFBF}, due to the higher accuracy of MiLAC-aided LS relative to digital LS, as observed in Fig.~\ref{fig.CE_NMSE}. The sum rate gain is significant for $G = 2$, where MiLAC-aided \ac{R-ZFBF} with LS-estimated CSI nearly matches digital \ac{R-ZFBF} with MMSE-estimated CSI. For $G=8$, however, the gain becomes smaller, and MiLAC-aided \ac{R-ZFBF} with LS-estimated CSI approaches digital \ac{R-ZFBF} with LS-estimated CSI.

\begin{figure}[!t]
\centering
\subfloat[$G = 2$]{\includegraphics[width=0.495\columnwidth]{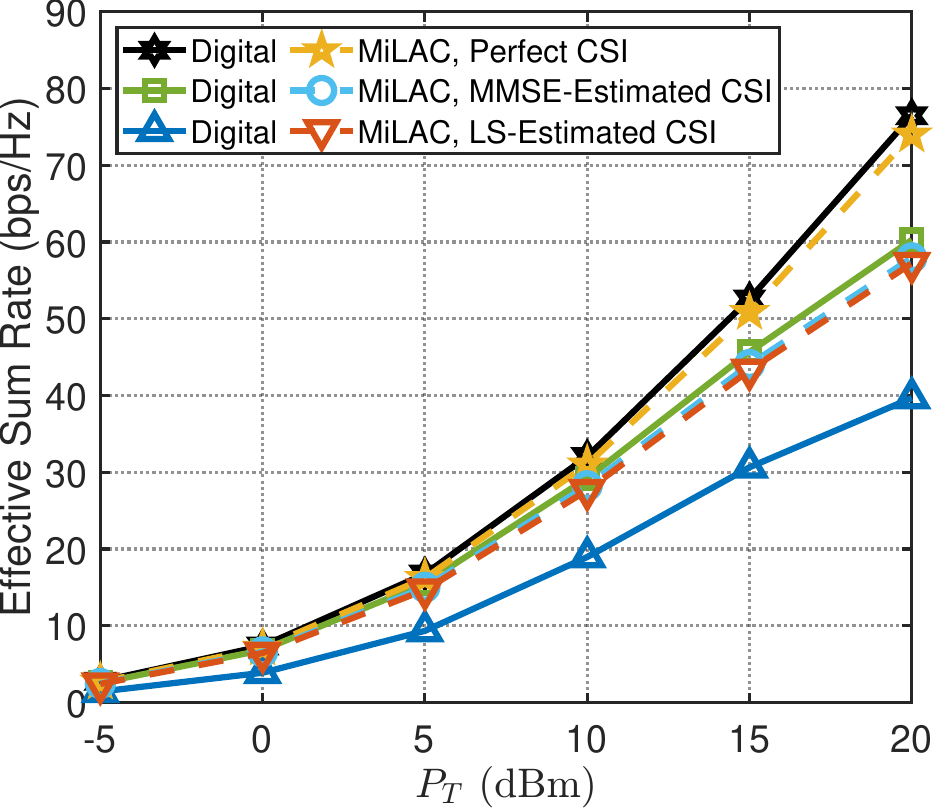}
\label{fig.BF_effective_sum_rate.a}}
\subfloat[$G = 8$]{\includegraphics[width=0.495\columnwidth]{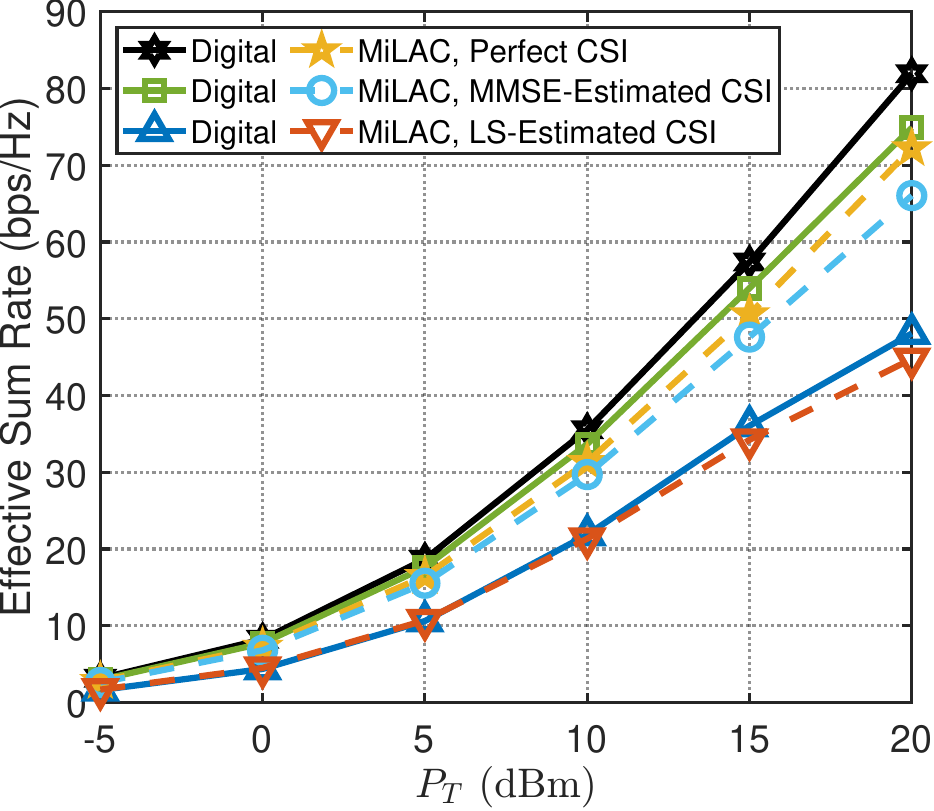}
\label{fig.BF_effective_sum_rate.b}}
\caption{Comparison of digital and MiLAC-aided beamforming: effective sum rate versus $P_T$.}
\label{fig.BF_effective_sum_rate}
\end{figure}

To further examine the impact of training overhead on data transmission, we extend the results in Fig.~\ref{fig.BF_sum_rate} by incorporating the scaling factor $(1-T_u/T)$ into the sum rate calculation. Specifically, as discussed in Remark~\ref{Remark 4}, we apply the scaling factors $(1-K\lceil r/K\rceil/T)$ and $(1-K\lceil \tilde{r}/K\rceil/T)$ to MiLAC-aided \ac{R-ZFBF} based on $\mathbf{H}_v$ for $G = 2$ and $\tilde{\mathbf{H}}_v$ for $G = 8$, respectively, while $(1-K/T)$ is applied to digital \ac{R-ZFBF} as the baseline. The resulting effective sum rate is plotted in Fig.~\ref{fig.BF_effective_sum_rate}, from which we make the following observations.
\textit{First}, when $G = 2$, MiLAC-aided \ac{R-ZFBF} suffers only a small performance loss relative to the digital baselines. This is because $r$ remains comparable to $K$ in the small-$G$ regime, leading to limited extra training overhead. 
\textit{Second}, when $G = 8$, the performance loss of MiLAC-aided \ac{R-ZFBF} relative to the digital baselines becomes more pronounced, because $\tilde{r}$ is larger relative to $K$ in the large-$G$ regime. In particular, MiLAC-aided \ac{R-ZFBF} with LS-estimated CSI achieves a slightly lower effective sum rate than digital \ac{R-ZFBF} with LS-estimated CSI. This indicates that, in this case, the loss caused by the extra training overhead outweighs the gain brought by the higher estimation accuracy of MiLAC-aided LS relative to digital LS. 
Overall, these observations reflect a fundamental trade-off in MiLAC-aided systems: employing only $L=K$ RF chains significantly reduces hardware complexity compared to fully digital systems, but necessitates extra training overhead, especially in the large-$G$ regime, thereby reducing the effective sum rate. Such a trade-off is well-documented in RF-chain-limited systems, e.g., hybrid digital-analog systems \cite{training_overhead_trade_off,CE_hybrid_classical}.

\begin{figure}[!t]
\centering
\subfloat[Small-$G$ Regime]{\includegraphics[width=0.495\columnwidth]{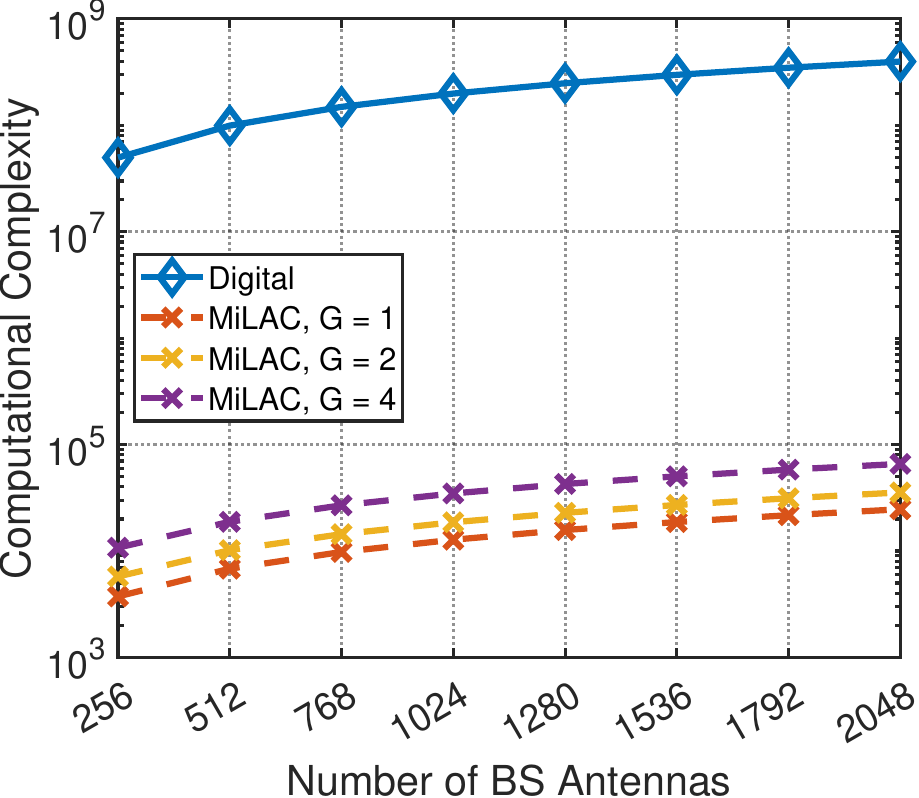}
\label{fig.BF_complexity.a}}
\subfloat[Large-$G$ Regime]{\includegraphics[width=0.495\columnwidth]{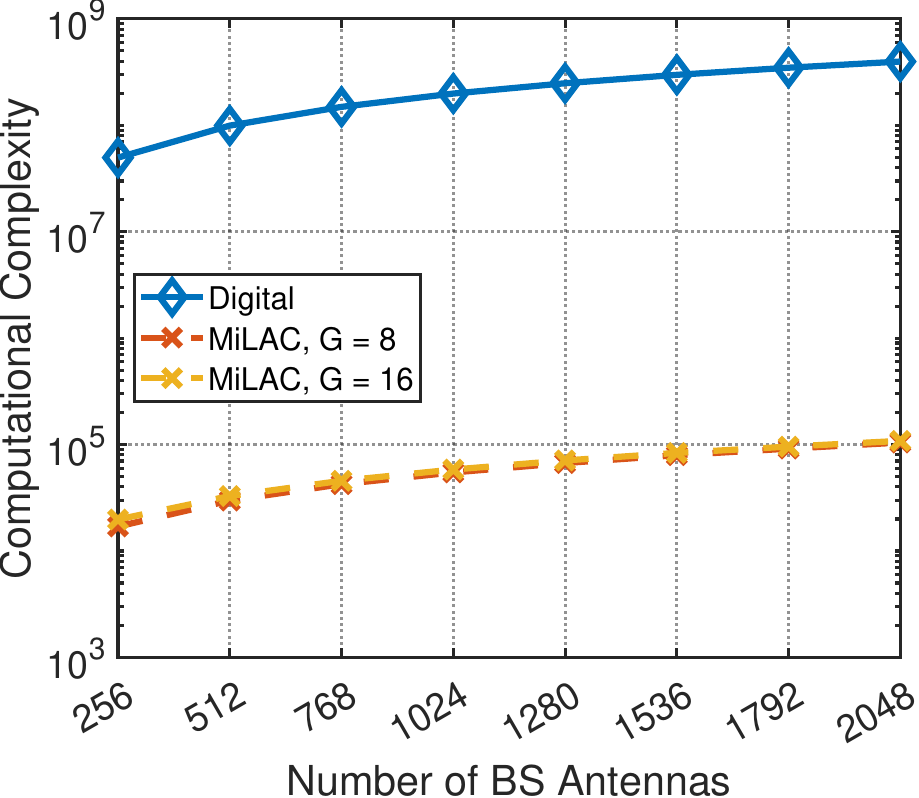}
\label{fig.BF_complexity.b}}
\caption{Comparison of digital and MiLAC-aided beamforming: computational complexity versus the number of BS antennas.}
\label{fig.BF_complexity}
\end{figure}

Finally, we evaluate the computational complexity of the proposed MiLAC-aided beamforming schemes versus the number of \ac{BS} antennas in Fig.~\ref{fig.BF_complexity}, with digital \ac{R-ZFBF} included as the baseline. 
From Fig.~\ref{fig.BF_complexity}, we have the following observations. 
\textit{First}, MiLAC-aided \ac{R-ZFBF} based on either $\mathbf{H}_v$ in the small-$G$ regime or $\tilde{\mathbf{H}}_v$ in the large-$G$ regime achieves much lower computational complexity than digital \ac{R-ZFBF}. The maximum reduction reaches $16108\times$ for $G=1$, corresponding to a saving of $3.97\times10^8$ real operations. 
\textit{Second}, in the small-$G$ regime, the computational complexity of MiLAC-aided \ac{R-ZFBF} increases with $G$, since its complexity scales as $6rK$ where $r$ rises with $G$. In contrast, in the large-$G$ regime, the complexity is almost insensitive to $G$, even though it scales as $6\tilde{r}K$. This indicates that $\tilde{r}$ exhibits little variation with $G$ in this regime, as also observed in Fig.~\ref{fig.CE_complexity.b}.

\section{Conclusion}
\label{Sec.6}
In this paper, we propose computationally efficient channel estimation schemes for \ac{MiLAC}-aided \ac{MU-MISO} systems with a limited number of \ac{RF} chains. 
Considering users partitioned into $G$ groups with different correlation matrices, we develop virtual channel estimation for small $G$ and global virtual channel estimation for large $G$.
Analytical results show that the proposed schemes achieve lower \ac{LS} \ac{MSE} and the same \ac{MMSE} \ac{MSE} as digital baselines, while greatly reducing computational complexity, at the cost of extra training overhead.

Based on low-dimensional channel estimates, we further develop \ac{MiLAC}-aided \ac{R-ZFBF} schemes. 
A cascade \ac{MiLAC} is introduced to realize the \ac{R-ZFBF} precoder directly from the virtual channel for small $G$ or the global virtual channel for large $G$, without reconstructing the full-dimensional channel.
Compared with digital \ac{R-ZFBF}, the proposed schemes are analytically shown to achieve the same sum rate with greatly reduced computational complexity, while incurring a minor sum-rate loss when uplink training overhead is considered.

Finally, numerical results validate the proposed schemes, confirming the predicted estimation and beamforming performance with substantial computational savings.
Particularly, the proposed schemes reduce computational complexity by up to $1540\times$ for channel estimation and $16108\times$ for beamforming. 
When training overhead is considered, a trade-off between the hardware complexity and the effective sum rate is observed.

In this work, we consider a fully-reconfigurable \ac{MiLAC} to characterize the fundamental performance and complexity limits of \ac{MiLAC}-aided channel estimation and beamforming. Extending the proposed framework to lossless and reciprocal \ac{MiLAC}s remains an important direction for future research.

\appendices

\section{Distribution of Group-Wise Noise}
\label{Ap.noise}

Recall that $\mathbf{n}_t  \sim \mathcal{CN}(\mathbf{0},\sigma^2\mathbf{I}_M) $ is \ac{AWGN}, independent over $t \in \mathcal{T}_u$, and $\mathbf{N}^{(f)}=[\mathbf{n}_{(f-1)L+1},\dots,\mathbf{n}_{fL}] \in \mathbb{C}^{M\times L}$ is the noise in frame $f$, $\forall f \in \mathcal{F}$. For a given group $g \in \mathcal{G}$, let
$\mathbf{A}_g^{(f)} \in \mathbb{C}^{\ell_g^{(f)} \times M}$
collect the rows of $\mathbf{G}^{(f)}$ assigned to group $g$ in frame $f$, where $0 \leq \ell_g^{(f)} \leq L$. In our setting, the rows of
$\mathbf{A}_g^{(f)}$ are selected from $\mathbf{U}_g^H$ and are
orthonormal, i.e.,
\begin{equation}
    \label{eq.Ap.noise.Ag_orthonormal}
    \mathbf{A}_g^{(f)}\big(\mathbf{A}_g^{(f)}\big)^H
    = \mathbf{I}_{\ell_g^{(f)}}.
\end{equation}
The contribution of frame $f$ to the $g$-th group effective noise is given by $\mathbf{A}_g^{(f)}\mathbf{N}^{(f)}$. By stacking all non-empty contributions across frames, we have the the $g$-th group effective noise 
\begin{equation}
    \mathbf{N}_g =
        \begin{bmatrix}
        \mathbf{A}_g^{(f_1)} \mathbf{N}^{(f_1)}\\
        \vdots\\
        \mathbf{A}_g^{(f_{F_g})} \mathbf{N}^{(f_{F_g})}
    \end{bmatrix}
    \in \mathbb{C}^{r_g \times L},
\end{equation}
where $\{f_1,\dots,f_{F_g}\}=\{f \in \mathcal{F}\;|\;\ell_g^{(f)}>0\}$ and $\sum_{j=1}^{F_g}\ell_g^{(f_j)} = r_g$. Equivalently, $\mathbf{N}_g=[\mathbf{n}_{g,1},\dots,\mathbf{n}_{g,L}]$, where the $l$-th column, $\forall l \in \mathcal{L}$, is
\begin{equation}
    \mathbf{n}_{g,l} =
    \begin{bmatrix}
        \mathbf{A}_g^{(f_1)} \mathbf{n}_{(f_1-1)L+l}\\
        \vdots\\
        \mathbf{A}_g^{(f_{F_g})} \mathbf{n}_{(f_{F_g}-1)L+l}
    \end{bmatrix}
    \in \mathbb{C}^{r_g \times 1}.
\end{equation}
Given $\mathbf{n}_t \sim \mathcal{CN}(\mathbf{0},\sigma^2\mathbf{I}_M)$, independent over
$t \in \mathcal{T}_u$, and \eqref{eq.Ap.noise.Ag_orthonormal}, we obtain
\begin{equation}
    \mathbb{E}\left[\mathbf{n}_{g,l}\mathbf{n}_{g,l}^H\right]
    = \sigma^2 \mathbf{I}_{r_g}.
\end{equation}
Moreover, $\mathbf{n}_{g,l}$ and $\mathbf{n}_{g,m}$ are independent, $\forall l,m \in \mathcal{L}, l \neq m$. Hence the entries of $\mathbf{N}_g$ are \ac{i.i.d.} with $\mathcal{CN}(0,\sigma^2)$, equivalently
\begin{equation}
    \mathrm{vec}(\mathbf{N}_g)
    \sim \mathcal{CN}(\mathbf{0},\sigma^2\mathbf{I}_{r_g L}).
\end{equation}

\section{Equivalence of MMSE MSE}
\label{Ap.MMSE_MSE_equivalence}

We first show that the MMSE MSE of estimating $\mathbf{h}_k$ in \eqref{eq.CE.digital_hk_MMSE_MSE} is identical to that of estimating $\mathbf{h}_{v,k}$ in \eqref{eq.CE.hvk_MMSE_MSE_full}. Since $\mathbf{R}_k=\mathbf{U}_g\mathbf{R}_{v,k}\mathbf{U}_g^H$, substituting this into \eqref{eq.CE.digital_hk_MMSE_MSE} yields
\begin{align}
    & \mathbb{E}\{\|\mathbf{h}_{k}-\mathbf{h}_k^{\text{MMSE}}\|^2\}\vphantom{\Big(\Big)^{-1}}\\
    & = \mathrm{tr}\Bigl(\mathbf{U}_g\mathbf{R}_{v,k}\mathbf{U}_g^H - p_kT_u\mathbf{U}_g\mathbf{R}_{v,k}\vphantom{\Big(\Big)^{-1}}\\
    & \quad\quad\;\;\times \mathbf{U}_g^H \bigl(p_kT_u\mathbf{U}_g\mathbf{R}_{v,k}\mathbf{U}_g^H+\sigma^2\mathbf{I}_{M}\bigr)^{-1}\mathbf{U}_g \vphantom{\Big(\Big)^{-1}}\label{eq.Ap.MSE_hk_intermsof_Rvk}\\
    &\quad\quad\;\;\times \mathbf{R}_{v,k}\mathbf{U}_g^H\Bigr)\vphantom{\Big(\Big)^{-1}}.
\end{align}
Moreover, using $\mathbf{U}_g^H\mathbf{U}_g=\mathbf{I}_{r_g}$, we have
\begin{align}
    &\mathbf{U}_g^H \big(p_kT_u\mathbf{U}_g\mathbf{R}_{v,k}\mathbf{U}_g^H+\sigma^2\mathbf{I}_{M}\big)^{-1}\mathbf{U}_g \\
    &= \bigl(p_kT_u\mathbf R_{v,k}+\sigma^2\mathbf I_{r_g}\bigr)^{-1}.
\end{align}
Substituting this into \eqref{eq.Ap.MSE_hk_intermsof_Rvk} and using the cyclic property of the trace, we obtain
\begin{align}
    & \mathbb{E}\{\|\mathbf{h}_{k}-\mathbf{h}_k^{\text{MMSE}}\|^2\}\vphantom{\Big(\Big)^{-1}} \\
    & = \mathrm{tr}\Big(\mathbf{R}_{v,k}- 
    p_kT_u\mathbf{R}_{v,k}\bigl(p_kT_u\mathbf R_{v,k}+\sigma^2\mathbf I_{r_g}\bigr)^{-1}\mathbf{R}_{v,k}\Big),
\end{align}
which is exactly the same as the MMSE MSE of estimating $\mathbf{h}_{v,k}$ in \eqref{eq.CE.hvk_MMSE_MSE_full}, given $L = T_u = K$. 
Similarly, \eqref{eq.CE.digital_hk_MMSE_MSE} can also be shown to be identical to the MMSE MSE of estimating $\tilde{\mathbf{h}}_{v,k}$ in \eqref{eq.CE.hvktilde_MMSE_MSE}, since $\mathbf{h}_k=\tilde{\mathbf{U}}\tilde{\mathbf{h}}_{v,k}$ and $\tilde{\mathbf{U}}^H\tilde{\mathbf{U}} = \mathbf{I}_{\tilde{r}}$. Consequently,
\begin{align}
    &\mathbb{E}\{\|\mathbf{h}_{k}-\mathbf{h}_k^{\text{MMSE}}\|^2\}\\
    &= \mathbb{E}\{\|\mathbf{h}_{v,k}-\mathbf{h}_{v,k}^{\text{MMSE}}\|^2\} =\mathbb{E}\{\|\tilde{\mathbf{h}}_{v,k}-\tilde{\mathbf{h}}_{v,k}^{\text{MMSE}}\|^2\}.
\end{align}

\bibliographystyle{IEEEtran}
\bibliography{refs}

\vfill

\end{document}